\documentclass[prb, preprint,showpacs]{revtex4}
\usepackage{epsfig,multirow}

\begin{document}
\title{Pinning of electron densities
in quantum rings by defects:
symmetry constraints and distribution of persistent currents}
\author{T. Chwiej} \affiliation{Faculty of Physics and Applied
Computer Science, AGH University of Science and Technology,\\ al.
Mickiewicza 30, 30-059 Krak\'ow, Poland}
\author{B. Szafran}
\affiliation{Faculty of Physics and Applied Computer Science, AGH
University of Science and Technology,\\ al. Mickiewicza 30, 30-059
Krak\'ow, Poland}

\begin{abstract}
We study the pinning of few-electron charge densities by weak perturbations to circular quantum
ring potentials using an exact diagonalization approach. The pinning results
in formation of single electron charge density islands distributed like classical particles.
We find that the pinning by weak defects is only allowed when the symmetry of the classical few-electron lowest-energy
configuration agrees with the symmetry of the external potential.
We indicate that whenever the pinning is allowed by the symmetry, its strength is an oscillatory function of the external magnetic field.
In the magnetic fields for which the pinning is maximal the dipole moment generated by the persistent currents changes
orientation from antiparallel to parallel to the external field in a continuous manner.
For confinement potentials of a higher symmetry than the one of a classical Wigner molecule the pinning is forbidden
and a discontinuous abrupt inversion of the dipole moments is observed.
When the pinning of single-electron islands is absent or weak the current distribution resembles
the one of a circular ring. For the maximal pinning instead of current loops running around the entire ring one observes formation of multiple current
vortices circulating around each  single-electron density island. We study the magnetic field generated by persistent currents
and find that at the dipole moment reversal the currents tend to screen the external field
in the region occupied by electrons. In consequence the magnetic field generated by the currents in the maximally pinned electron systems maps the
charge distribution. We also argue that the maximal pinning of charge densities is associated with smooth extrema of the chemical potentials
that for highly symmetric potentials -- in which the pinning is excluded -- are replaced by cusps.
\end{abstract}
 \maketitle
\section{Introduction}

Semiconductor quantum rings of nanometer size with both open and closed geometry
are produced by several fabrication techniques including self-assembly,\cite{lorke} surface
oxidation\cite{so} and etching techniques.\cite{hawrylak} Open quantum rings\cite{buti} attached to the electron reservoirs
by contacts are studied in the transport measurements \cite{transport} while closed rings are studied in context of the
optical and single-electron charging properties.\cite{lorke} When placed in an external magnetic field the
electron systems confined in quantum rings give rise to persistent currents which generate magnetic field on their own.\cite{rev} The persistent currents in closed nanorings were subject of a number of theoretical papers\cite{rev}
but the experimental detection of the magnetization signal came only relatively recently,
\cite{fomin} although
measurements on a single mesoscopic ring have a longer \cite{mm2s} history.
The self-assembled rings grow in an elliptic geometry.\cite{foomin} The oxidized and
etched rings may have an arbitrary shape including the circular one. Nominally
circular rings produced by these techniques are bound
to contain defects due to interface roughness in particular. Even weak defects to the circular
potential should influence the properties of the electrons confined in a quantum ring due to the reentrant
ground-state degeneracy appearing at the Aharonov-Bohm ground-state angular momentum transitions.\cite{rev}
A competition of the localization at the defects and extended electron states is
likely to appear. The magnetic dipole moment generated by persistent currents
should significantly depend on the form of electron localization. In particular the
persistent current loops circulating around the ring can be broken by formation
of localized states. The purpose of the present paper is to determine the
distribution of the persistent currents and the resulting magnetic properties of a few-electron
system in a circular quantum ring perturbed by weak defects.

Electrons confined in semiconductor nanostructures exhibit strong
correlated properties  when the electron-electron interaction energy
exceeds the single-particle quantization energies. This occurs for
instance in large quantum dots \cite{lqd} containing a small number
of electrons or at high magnetic field \cite{mani} for fractional
filling of the lowest Landau level. Favorable conditions for
appearance of strong electron-electron correlation exist also in
quantum rings \cite{rev} in which the energy levels are nearly
degenerate with respect to the angular momentum. This near
degeneracy is a counterpart of the degeneracy of Fock-Darwin energy
levels \cite{mani} with the lowest Landau level which occurs at high
magnetic field ($B$) in quantum dots. In circular quantum dots the
ground-state of a few-electron system is an intermediate
\cite{jain,tavernier} phase between the electron liquid and the
electron solid with a dominant character of the latter at high
magnetic fields. Since in quantum rings the near degeneracy with
respect to the angular momentum is observed also at zero magnetic
field, the electron (Wigner) crystal can be formed also for $B=0$
provided that the electron density is low. \cite{mani,rev,sigmund} A
natural distinction between the electron liquid and solid phases is
the range of the electron-electron correlation observed in the
pair-correlation function.\cite{szafranchwiej} This criterium is
however of a limited use for finite systems, and in particular for
few-electron quantum dots in which one can hardly speak of the long
range order. A more useful criterium allowing to distinguish a
quantum liquid from the quantum solid is the reaction of the system
to a weak external perturbation. The electron crystal, in contrast
to the electron liquid, is expected to be pinned,\cite{glazman} i.e.
extracted from the internal degrees of freedom \cite{maksym,mani} of
the system to the laboratory frame, by an arbitrary weak external
perturbation.

In this paper we consider
 pinning of the electron systems by
weak and short range perturbations to the circular quantum
ring potential in two dimensions. Due to the pinning, the
electron correlation appears in a form of crystallized single-electron charge density
islands occupying positions. We demonstrate that the
crystallization of the electron density is occasionally prohibited
by disagreement of the symmetry of the perturbed ring potential with
the natural, i.e. classical,\cite{peeters} symmetry of a
few-electron Wigner molecule. We find that when the Wigner
crystallization is allowed by the symmetry, the strength of the
localization of the single-electron islands strongly oscillates with
the external magnetic field. The most pronounced pinning of electron
densities occurs at the local smooth energy maxima of the
ground-state energy that appear due to opening of avoided crossings
in the energy spectra resulting of the angular momentum eigenstates mixing by the potential of the defect.

We find that the crystallization of the electron density has its
clear signatures in the properties of the system which are
accessible to the experiment. We calculate the magnetic dipole
moment produced by the persistent currents as well as the chemical
potentials of the confined systems.  We also consider the chemical potential of the ring
confined systems. Rings embedded in charge-tunable structures
\cite{lorke} are occupied by subsequent electrons for gate voltages and
external magnetic fields which align the chemical potential of the confined system
with the Fermi level of the electron reservoir. The ground-state transitions
of the spatial and spin symmetry results in occurrence of cusps in
the charging lines in function of the magnetic field. We show that
the maximal pinning of the charge density is accompanied by
replacement of the cusps by smooth extrema.

We  consider systems with up to $N=3$ electrons and solve the
effective mass equations for the system confined within a
two-dimensional plane with the external perpendicular magnetic field
using an exact diagonalization approach. The calculations presented
below are performed for etched InGaAs/GaAs quantum rings
\cite{hawrylak} which are produced with an intentionally designed
shape. We study rings with a single defect placed on a circumference
of the ring as well as rings with two defects placed symmetrically
with respect to the center of the ring [see Fig. 1]. The results that we obtain
for symmetrically placed defects are also qualitatively relevant for
self-assembled InGaAs quantum rings which are known to have an
elliptical confinement potential.\cite{fomin,pcct4,foomin} For a
single defect an appearance of $N$
single-electron islands is observed for any electron number. In the
other case -- for which the symmetry is lowered from circular to
elliptic -- the charge density with $N$ islands is only observed for
$N=2$. The elliptic potential is invariant with respect to ${\bf r}\rightarrow - {\bf r}$ operation,
but for the odd $N$ the classical Wigner molecules
do not have inversion symmetry. In consequence the pinning of
single-electron islands is not observed for odd $N$.

We determine the magnetization generated by the persistent currents.\cite{pcct0}
The orientation of the magnetic dipole moment generated by the currents oscillates
between paramagnetic and diamagnetic with the flux of the external field.
For circular quantum rings when $B$ increases the transitions
of the dipole moment from paramagnetic to diamagnetic occurs in a
continuous manner at the ground-state energy minima. On the other
hand the reversals of the dipole moment from diamagnetic to
paramagnetic orientation are abrupt and occur at the ground state
symmetry transformations associated with the Aharonov-Bohm
ground-state angular momentum transitions. On the contrary for
systems in which the natural symmetry of the Wigner molecule agrees
with the symmetry of the confinement potential this dipole moment
reversal is continuous. Moreover,  for magnetic fields corresponding
to the strongest pinning the diamagnetic and paramagnetic
contributions to the dipole moment are exactly equilibrated and
cancel each other. In one-dimensional models \cite{sigmund} the currents
circulating around the ring are stopped when the pinned charge
density takes the form of a Wigner crystal.
We find that although at the maximal charge density pinning the
current loop around the ring is indeed broken, the currents are actually not stopped but
form separated vortices rotating around the
single-electron density islands.
When the
pinning of the density is absent or weak one observes concentric
current loops running in opposite directions at the  inner and
outer edges of the ring like for an ideally circular ring. Although
the current distribution is very different in cases when the pinning
is minimal and maximal, at the magnetic dipole reorientation in both
cases the currents tend to generate magnetic fields that screen the
external field within the electron charge density maxima.

\begin{figure}[ht!]
\centerline{\hbox{\epsfysize=50mm
               \epsfbox[114 327 642 635 ] {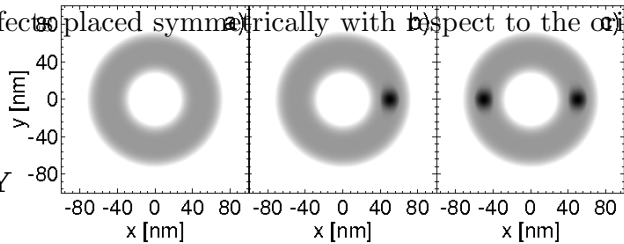}\hfill}}
\caption{Plots of the confinement potential of a clean
quantum ring (a), the ring with a single defect (b), and two defects
placed symmetrically with respect to the origin.}\label{pot}
\end{figure}

\section{Theory}
In both the self-assembled \cite{lorke} and the etched quantum rings\cite{hawrylak} the in-plane
dimensions of the rings exceed several times their height (the size in the growth direction). The electron-states in the growth direction
can be therefore treated as frozen by the quantum size effect, which usually justifies the two-dimensional
approximation in the studies of the confined systems. Moreover, for rings with circumference largely
exceeding the radial width of the ring, the reaction of electrons
to perturbations of the confinement potential and to their mutual interaction, in most cases
appears in the angular and not in the radial degree of freedom. For this reason many effects can be described within
strictly one-dimensional models. However, in this paper we also describe persistent currents that circulate around the single-electron charge density islands. Clearly, description of these
vortices is out of the reach of the one-dimensional models.  Therefore, we consider a two-dimensional
confinement potential
\begin{equation}
V(\vec{r})=-V_{0}\exp\bigg (-\bigg | \frac{|{\bf r}|-R_{0}}{\sigma_{0}}
\bigg |^\alpha \bigg )
\end{equation}
where ${R}_{0}$ denotes the mean radius of the ring, and $\sigma_0$ the radial width of the ring.
We assume $\alpha=20$ for which the potential is nearly a square quantum well.
The parameters of the rings are adopted for the etched In$_{0.1}$Ga$_{0.9}$As/GaAs quantum rings after Ref. [\cite{ostatni}]:
$V_{0}=50$ meV, $R_{0}=30$ nm,  $\sigma_{0}=20$ nm, the electron band effective mass
 $m^{*}=0.05m_0$ and the dielectric constant $\varepsilon=12.4$.  The plot of a clean quantum ring is given in Fig. \ref{pot}(a).
 As the confinement perturbations we introduce a single or two shallow Gaussian cavities $-V_{p}\exp\bigg (-\bigg | \frac{\vec{r}-\vec{R}_d}{\sigma_{d}}
\bigg|^2 \bigg )$, where $R_d$ is the location of the defect. The range of the perturbation was assumed $\sigma_d=10$ nm and its depth $V_p=0.5$ meV.
The potentials for the perturbed rings are plotted in Fig. \ref{pot}(b) and (c).

We consider a few-electron Hamiltonian
\begin{equation}
{H}=\sum_{i=1}^{N}{h}_{i}+\sum_{i=1,j>i}^{N}\frac{e^2}{4\pi\varepsilon_0\varepsilon r_{ij}}
\label{hn}
\end{equation}
where ${h}_{i}$ is single electron energy operator. In order to diagonalize the operator (\ref{hn}) we apply the configuration
interaction approach in which we look for the eigenvectors in a basis of
many electron wavefunctions with fixed values of total spin $S$ and its projection on the $z$ axis
$S_{z}$, which are generated by the projection operators \cite{low}
as linear combinations of Slater determinants
built of eigenfunctions of a single-electron Hamiltonian:
\begin{equation}
{h}_{i}=\frac{({\bf{p}}+e{\bf{A}}({\bf{r}}_{i}))^{2}}{2m^{*}}+V_{ext}
({\bf  r } _ { i } ).
\label{hi}
\end{equation}
where the vector potential is taken in the symmetric gauge: ${\bf {A}}({\bf{r}})=B(-y,x,0)/2$.
The single-electron eigenproblem is diagonalized in a basis
\begin{equation}
\phi_{i}({\bf{r}})=\sum_{\alpha=1}^{N}C_{\alpha}^{(i)}f_{\alpha}({\bf{r}})
\label{jedn},
\end{equation}
where the basis function are assumed in the form
\begin{eqnarray}
f_{\alpha}({\bf{r}})=\exp\bigg(-\frac{({\bf {r}}-{\bf{R}}_{\alpha})^2} {2\sigma^{2} } \bigg)
\exp\bigg(-\frac{ie}{2\hbar} ({\bf {B}}\times {\bf{R}}_{\alpha})\cdot {\bf {r}}
\bigg),
\label{fa}
\end{eqnarray}
where  ${\bf {R}}_{\alpha}=(x_{\alpha},y_{\alpha})$ is the center of the Gaussian,
 $\sigma$ describes its spatial extension and the imaginary exponent introduces the magnetic translation,
which ensured the gauge invariance of the basis.
The centers are distributed on a regular square mesh of $50\times 50$ points. The size of the mesh, i.e., the exact positions of the Gaussians
and the value of $\sigma$ parameter are selected variationally. Tests of the approach for a two-electron problem are given in detail
in Ref. [\cite{naszarx}].
In the precedent work on few-electron problems in semiconductor nanostructures in external magnetic field
sets of several Gaussian functions were used on several occasions.\cite{kainz,szafran,yann} Usually
the choice of the exact positions of the Gaussian centers involves a time-consuming non-linear optimization of the variational parameters
for each considered state and the magnetic field value. The mesh of 250 Gaussian functions which we apply here turns out to be extremely
flexible so that universally optimal non-linear parameters can be indicated.
The energy spectrum in a large magnetic field range can be quite accurately obtained for fixed values of the $\sigma$ parameter and the position of Gaussians.\cite{naszarx}

The persistent charge current densities are calculated as $\langle\Psi|{\bf j}_p+{\bf j}_d|\Psi\rangle$, where, $\Psi$ is the
$N$-electron eigenstate, and ${\bf j}_p$ and  ${\bf j}_d$ are the paramagnetic and diamagnetic current operators
defined as
\begin{equation}
{\bf j}_p=\frac{-e}{2m^*} \sum_{i=1}^N \left[-{\bf p} \delta({\bf r}-{\bf r}_i)+2\delta({\bf r}-{\bf r}_i){\bf p}_i\right],
\end{equation}
and
\begin{equation}
{\bf j}_d=\frac{-e}{2m^*} \sum_{i=1}^N \left[2e{\bf A}\delta({\bf r}-{\bf r}_i)\right],
\end{equation}
respectively.
The magnetic field generated by the persistent current is calculated from the Biot-Savart law
\begin{equation}
{\bf B}({\bf r})=\frac{\mu_0}{4\pi} \int \frac{{\bf j}\times\left({\bf r}-{\bf r}'\right)}{|{\bf r}-{\bf r'}|^3}d{\bf r}'.
\end{equation}
Deviation of the diamagnetic permeability of InAs/GaAs from unity is totally negligible. 
\section{Results}
\subsection{Clean circular ring}
\begin{figure}[ht!]
\centerline{\hbox{\epsfysize=45mm
               \epsfbox[39 250 458 610] {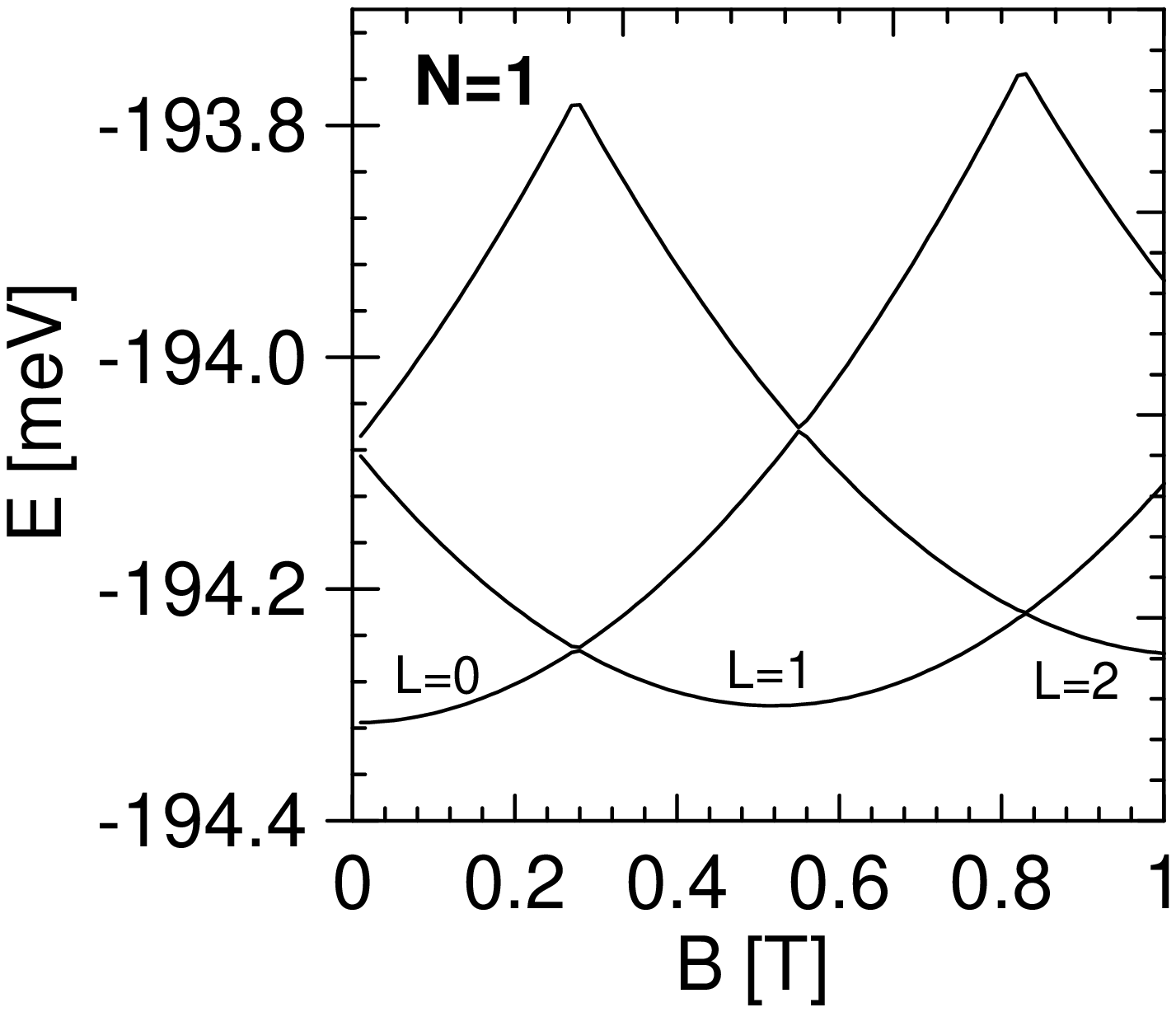}a)\epsfysize=45mm
               \epsfbox[39 250 458 610] {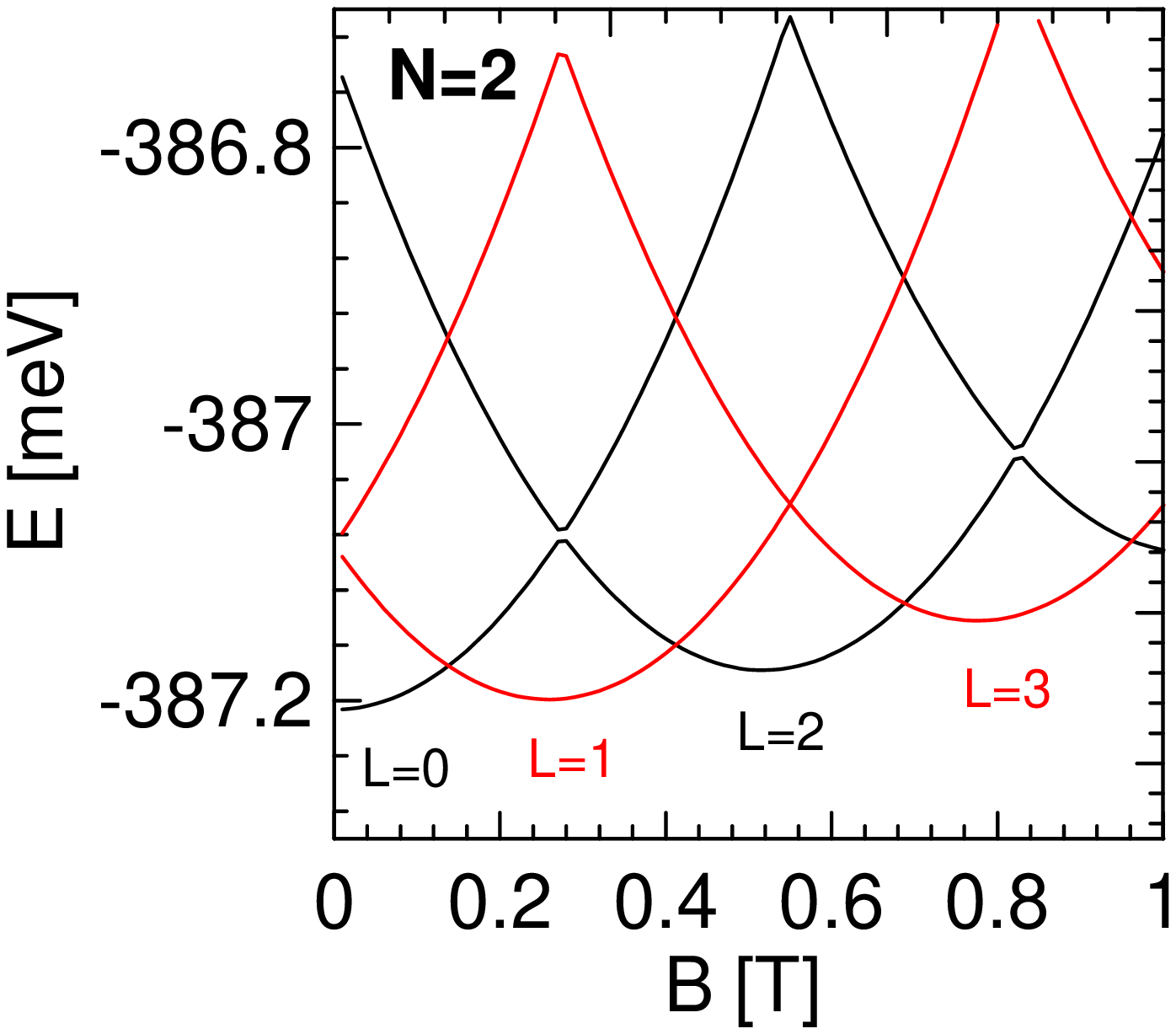}b)\epsfysize=45mm
               \epsfbox[39 250 458 610] {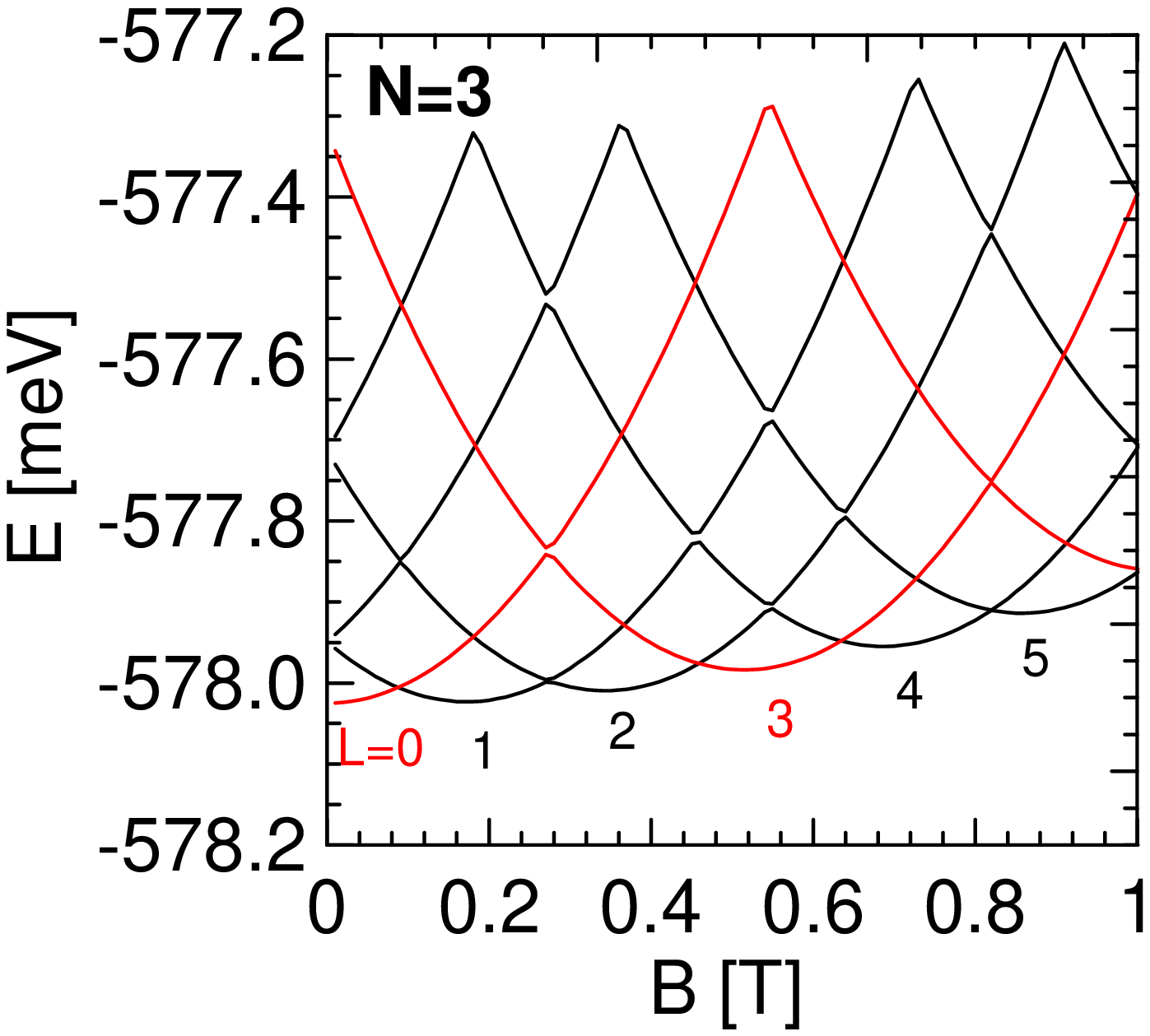}c)
               \hfill}   \vspace{0.5cm}
               }
               \centerline{\hbox{\epsfysize=45mm
               \epsfbox[39 250 458 610] {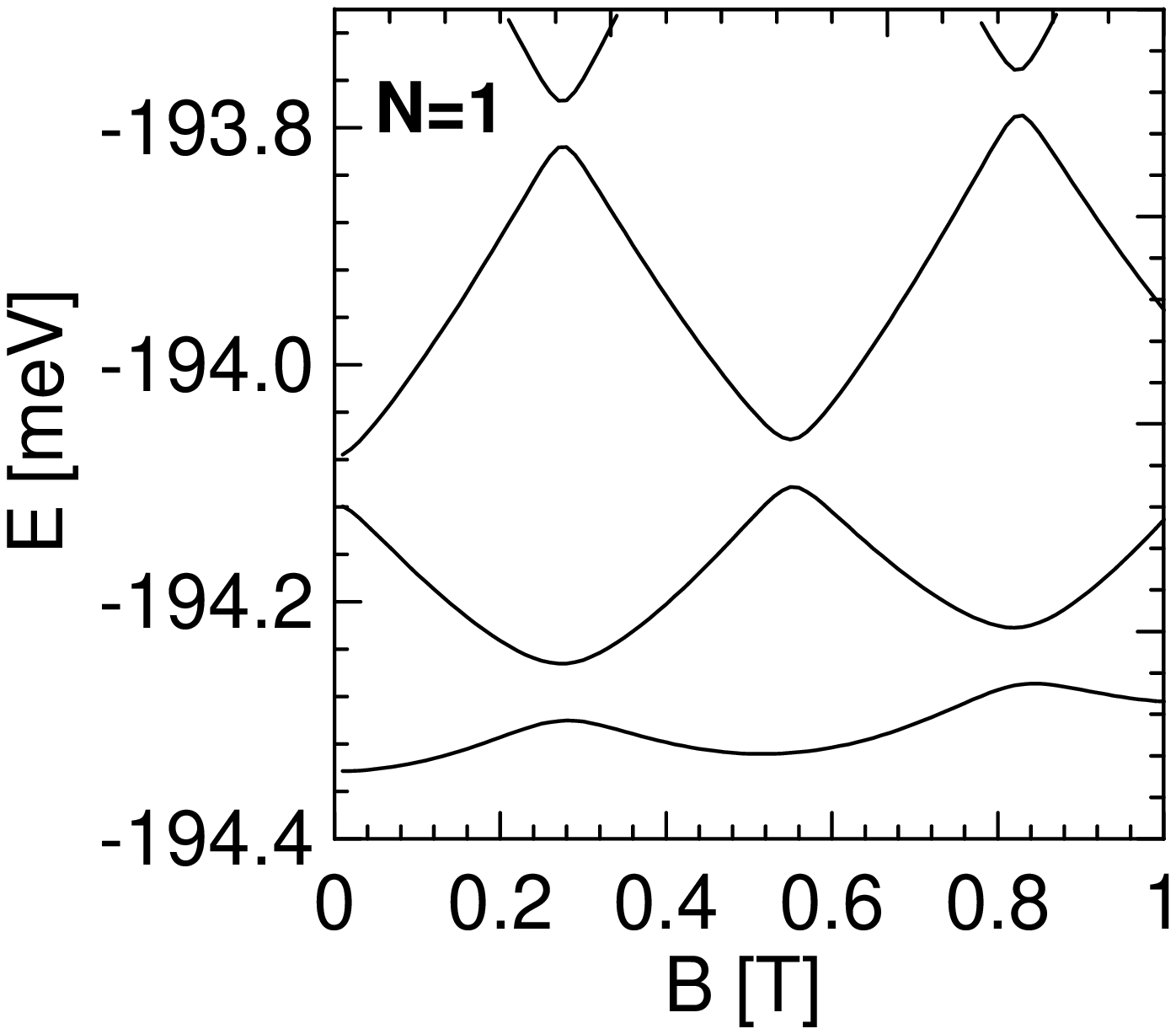}d)\epsfysize=45mm
               \epsfbox[39 250 458 610] {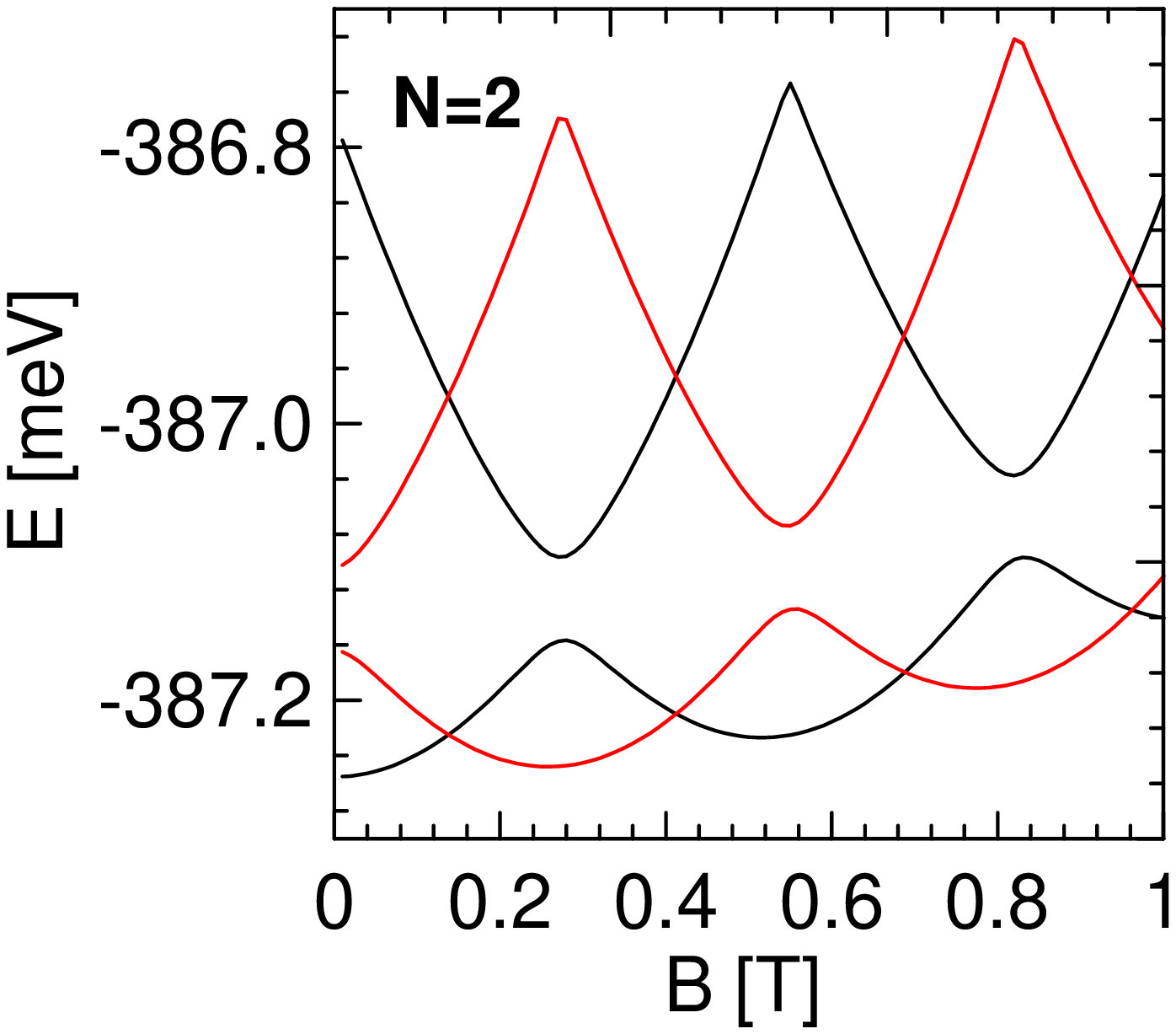}e)\epsfysize=45mm
               \epsfbox[39 250 458 610] {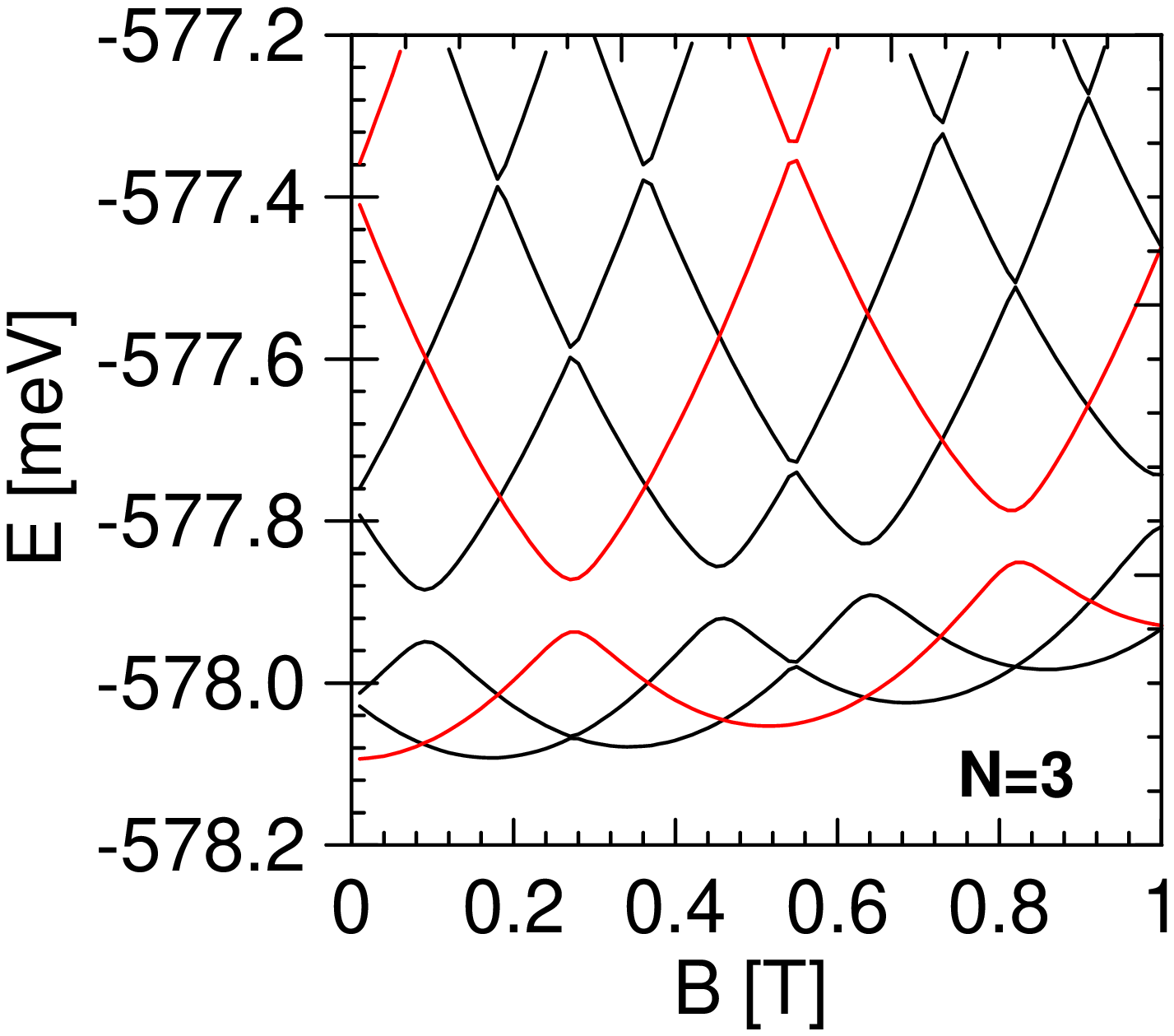}f)
               \hfill}\vspace{0.5cm}
               }
\centerline{\hbox{\epsfysize=45mm
               \epsfbox[39 250 458 610] {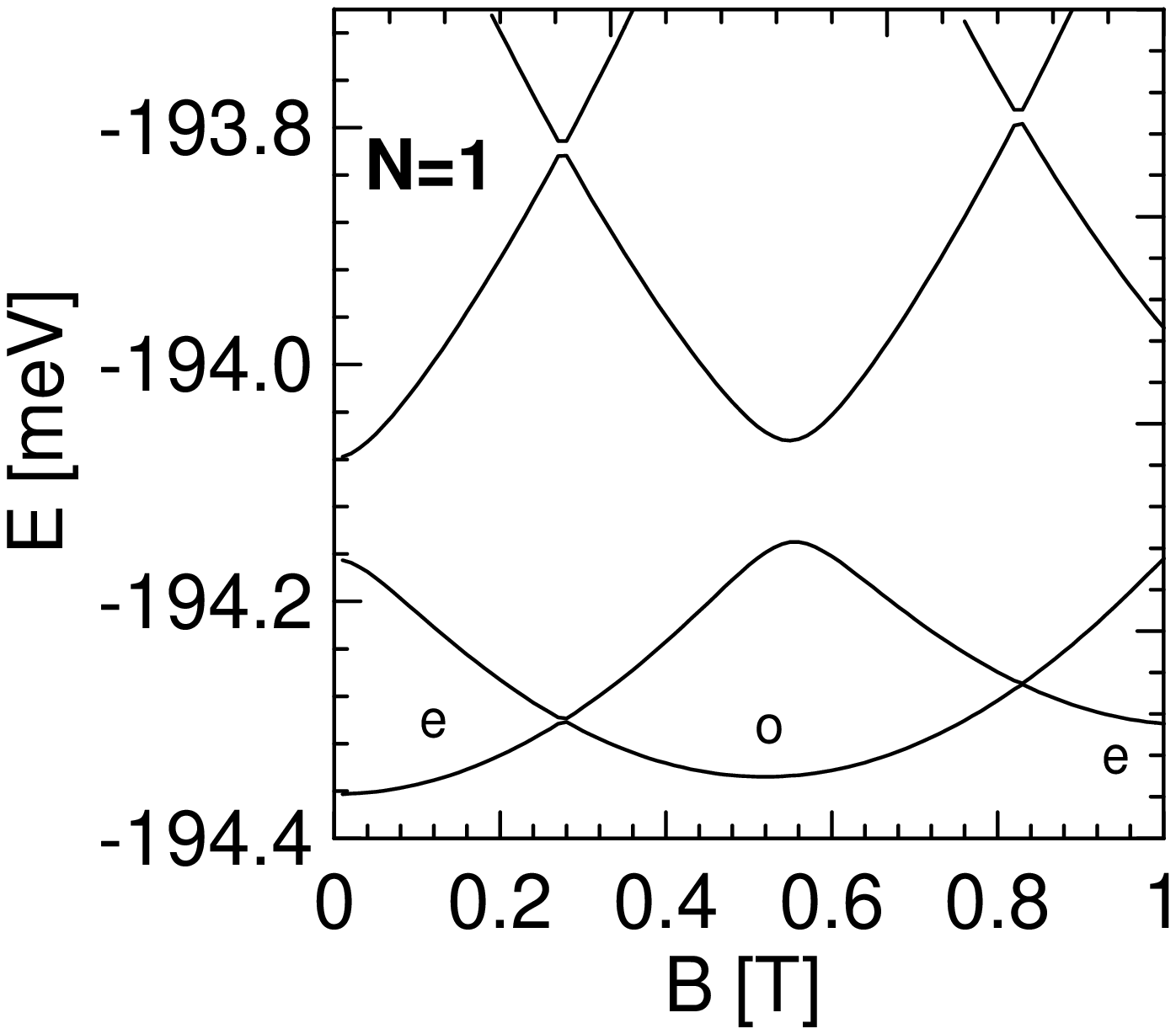}g)\epsfysize=45mm
               \epsfbox[39 250 458 610] {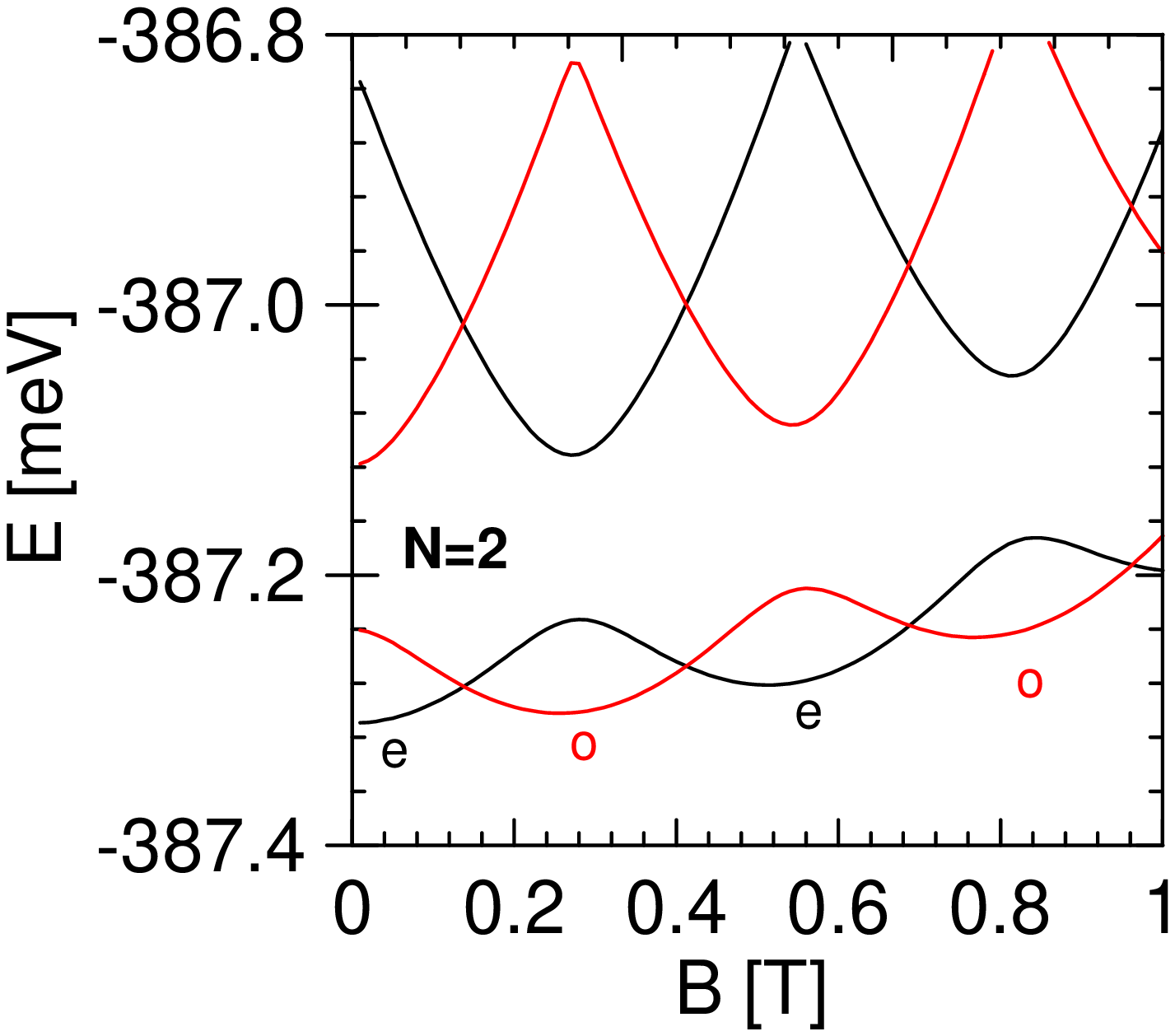}h)\epsfysize=45mm
               \epsfbox[39 250 458 610] {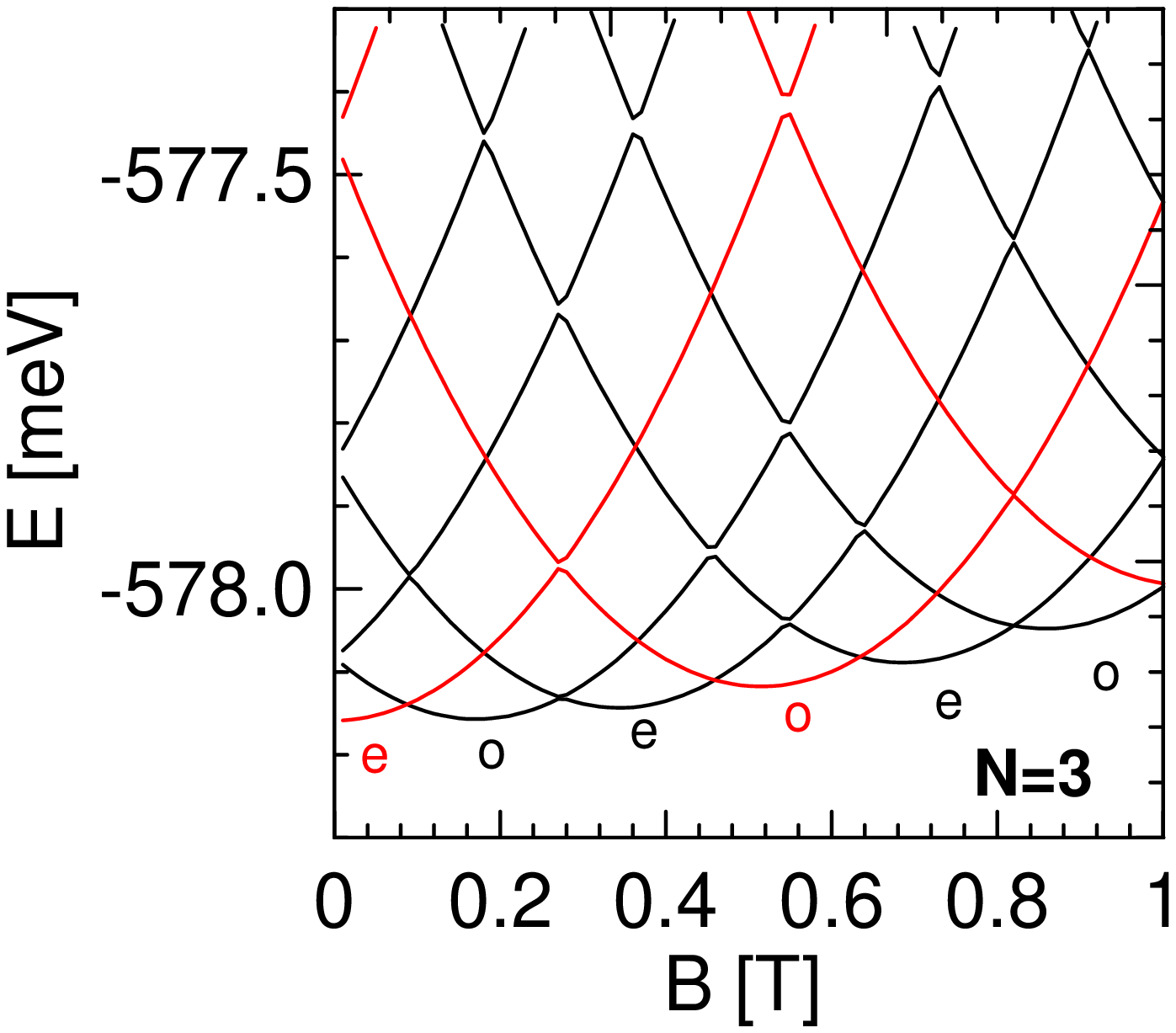}i)
               \hfill}
               }
\caption{Energy spectra for one (a,d,g) two (b,e,h) and three electrons (c,f,i)
in a pure quantum ring (a-c), with a single defect (d-f) and with two defects
(g-i) placed symmetrically with respect to the center of the ring. In plots for $N=2$ and $N=3$ energy
levels corresponding to the spin-polarized states are plotted with the red color.}
\label{spectra}
\end{figure}

Results for the circular ring are presented as the point of
reference to the perturbed rings.  The energy spectra for $N=1,2$
and 3 are presented in Fig. \ref{spectra}(a),(b) and (c),
respectively. In the ground-state we observe the angular momentum
transitions. The magnetic field range corresponding to the stability
of subsequent angular momentum states decreases with the number of
electrons, which is due to the fractional Aharonov-Bohm effect
\cite{fqhe} for the few-electron system. In particular, for $N=1$
the $L=1$ state is stable in the interval of magnetic field of
length 0.55 T. This value corresponds to the flux quantum passing
through a strictly 1D ring of radius 49 nm, which well agrees with
the mean radius value adopted in the present model. For $N=2$ and
$3$, the $L=1$ state is stable within an interval of length nearly
equal to 0.55T$/N$. For $N>1$ the angular momentum transitions are
accompanied by the spin transitions. For two electrons the
ground-states of odd $L$, and for three electrons the ground-states
of $L$ equal to integer multiples of 3, are spin polarized. The
energy levels corresponding to the spin-polarized $N=2$ and $N=3$
states are plotted in ref in Fig. \ref{spectra}.

\begin{figure}[ht!]
\centerline{\hbox{\epsfysize=150mm
               \epsfbox[39 63 491 754] {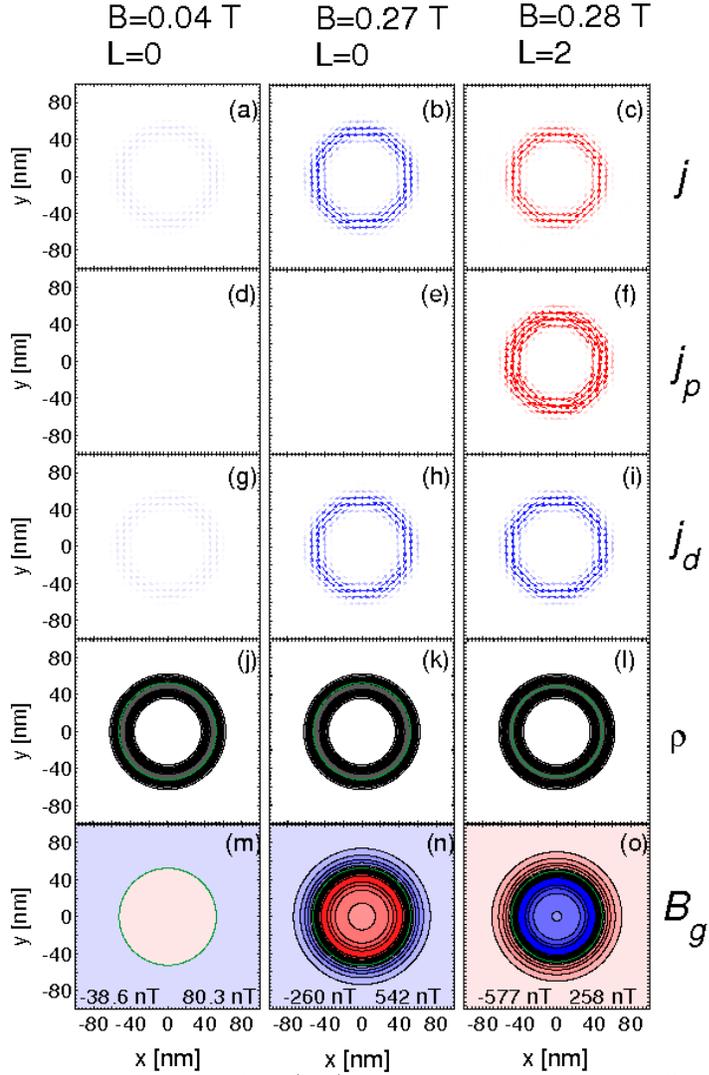}\hfill}}
\caption{Persistent current distribution (a-c), the paramagnetic
contribution to the current (d-f), the diamagnetic current (g-i), the
charge density (j-l) and the $z-$ component of the magnetic field
generated by the persistent currents $B_g$ calculated within the plane
of confinement $z=0$ for $N=2$ electrons in a circular ring. All the plots correspond
to the lowest energy singlet. The left and the central column of plots
correspond to $L=0$ and the right column to $L=2$.
The color and vector scales are the same in each row. For the currents
the same scale is applied in all the figures (a-h). Red vectors correspond
to positive (counterclockwise) angular orientation and the blue ones to negative (clockwise) orientations.
 In (j-o) the green
line shows the contour on which $B_g$ changes sign. In (m-o) red
color corresponds to positive $B_g$ (opposite to the external field orientation).
Negative $B_g$ corresponds to blue region.
In each of the plots (m-o) the color scale is the same we additionally give the
 minimal and maximal values of $B_g$ obtained for a given $B$ . \label{2ecsi}}
\end{figure}

For illustration of the distribution of the charge, current, and the magnetic field generated by the ring-confined electrons
we chose the case of two electrons (qualitatively identical results are obtained for
both $N=1$ and $N=3$). The persistent current distribution
for the lowest-energy
singlets is displayed in Fig. \ref{2ecsi}(a-c),
and the paramagnetic and diamagnetic contributions to the current in Fig. \ref{2ecsi}(d-i).
 At $B=0$ the two-electron ground-state
is the spin-singlet with zero angular momentum. In this state at $B=0$ both the paramagnetic
and diamagnetic currents are zero. For non-zero $B$, the paramagnetic current of the $L=0$ state
remains zero.\cite{odnosnikd}
For the magnetic field oriented antiparallel to the $z$-axis the diamagnetic current runs
counterclockwise around the ring. It generates the magnetic field $B_g$ which is parallel to the $z$-axis (i.e. opposite
to the external field) inside the ring [see Fig. \ref{2ecsi}(m,n)] and parallel to the $z$-axis outside the ring.
In figures illustrating the results of this paper we mark the regions of positive $B_g$ (opposite to the external field) with the red color,
and the regions in which
$B_g$ is parallel to $B$ by blue color [see. Fig. \ref{2ecsi}(m-o)].
In the plots of $B_g$ we list the most positive and most negative values of
the magnetic field generated by the persistent current.
The darker shade of the blue / red color the larger the absolute value of $B_g$.
Note that
since $\nabla \cdot B_g=0$ the surface integral of the magnetic field is strictly zero --
positive flux of $B_g$ within the ring (within the current loop) is equal to the negative flux of $B_g$ outside the ring.

For $B\simeq 0.275$ T the lowest-energy spin state changes its angular momentum from $L=0$ to $L=2$ [see the crossing
of the black lines in Fig. \ref{spectra}(b)].
The charge density of the nearly degenerate lowest-energy singlets for $L=0$ and $L=2$
is nearly identical [see Fig. \ref{2ecsi}(k) and Fig. \ref{2ecsi}(l)]. In consequence
the diamagnetic contribution to the current is also nearly the same for both the states
[cf. Fig. \ref{2ecsi}(b) and Fig. \ref{2ecsi}(c)]. Nevertheless, for $L=2$
the paramagnetic contribution to the current is roughly twice stronger
than the diamagnetic current [see Fig. \ref{2ecsi}(f)]
at the magnetic field corresponding to the level crossings of $L=0$ and $L=2$ lowest-energy
eigenstates.
The resulting persistent current has
a pure paramagnetic character [see Fig. \ref{2ecsi}(c)]. The paramagnetic current
circulates clockwise around the ring [cf. Fig. \ref{2ecsi}(c)]
and generates the magnetic field in the direction of the external field within the current loop
and opposite to the external field outside the loop [cf. Fig. \ref{2ecsi}(o)].
Note that the maximal and minimal values of $B_g$ for $L=2$ at $B=0.28$ T are
nearly inverse of the values obtained for $L=0$ at $B=0.27$ T [see the bounds listed
in Fig. \ref{2ecsi}(n,o)] .
In the charge density plots [Fig. \ref{2ecsi}(j-l)] we additionally marked the
contour in which $B_g$ changes orientation with the green line. For $L=0$ most of the electron charge
stays inside the region in which $B_g$ is opposite to $B$, but already for $L=2$ the
nodal $B_g$ line cuts the maximal electron density region in half. Both the densities
are nearly identical, it is the $B_g$ contour which decreases in size at the angular momentum transition.

\begin{figure}[ht!]
\centerline{\hbox{\epsfysize=150mm
               \epsfbox[18 133 588 774] {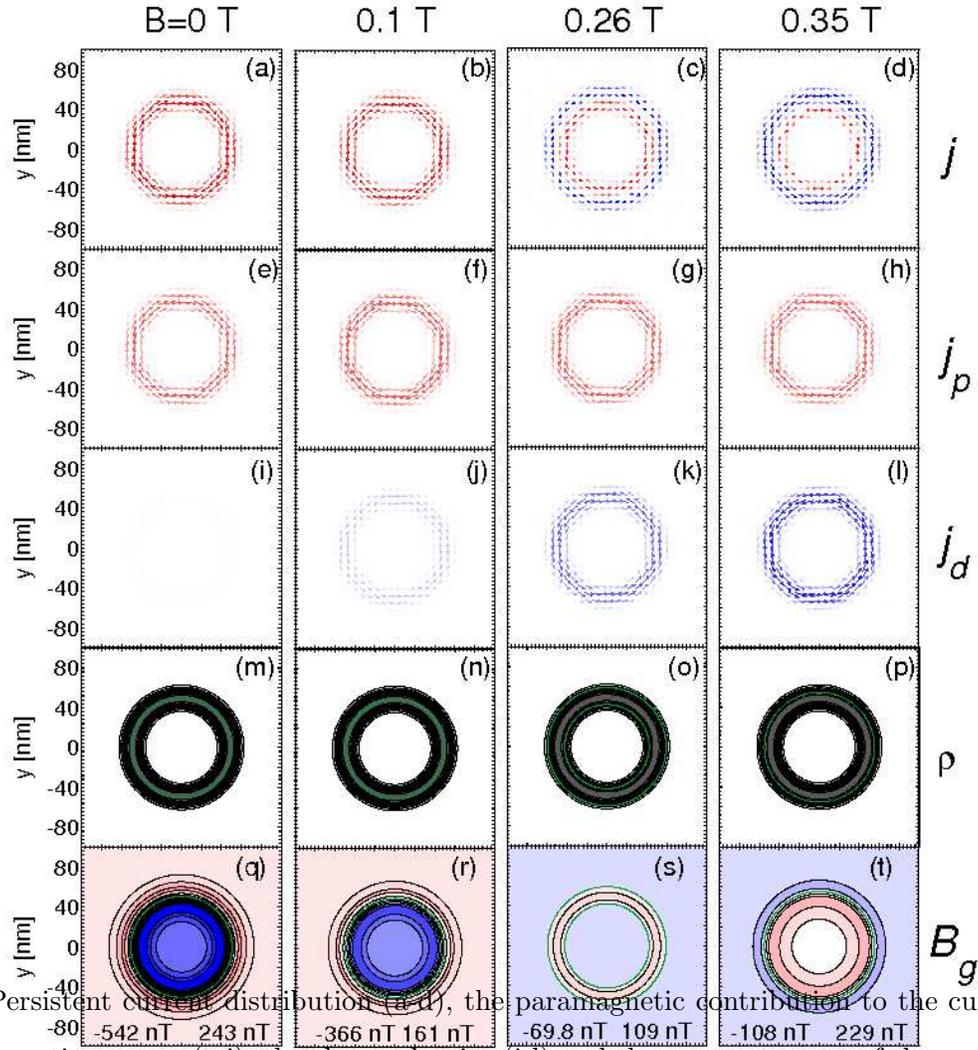}\hfill}}
\caption{Persistent current distribution (a-d), the paramagnetic
contribution to the current (d-f), the diamagnetic current (g-i), the
charge density (j-l) and the $z-$ component of the magnetic field
generated by the persistent currents $B_g$. All the plots correspond
to the lowest energy triplet with $L=1$.
The color and vector scales are the same in each row. In (i-t) the green
line shows the contour on which $B_g$ changes sign. In (q-t) the red
color corresponds to positive $B_g$ (opposite to the external field orientation).
Negative $B_g$ corresponds to blue region.
In each of the plots (q-t) the color scale is the same we additionally give the
 minimal and maximal values of $B_g$ obtained for a given $B$. \label{2ecti}}
\end{figure}

Fig. \ref{2ecsi} illustrates the effects related to the angular momentum transition
for a fixed spin state. The transition described in Fig. \ref{2ecsi} occurs outside of the ground-state [cf. the
crossing of singlet (black) energy levels in the first excited state near $B=0.275$ T in Fig. \ref{spectra}(b)].
Let us now concentrate on the evolution of the quantities calculated for a fixed value of $L$ when the magnetic field is increased.
We selected for this purpose the spin triplet of $L=1$, which is the two-electron ground state
near $B=0.25$ T [see Fig. \ref{spectra}(b)].
The distribution of the paramagnetic current weakly depends on $B$ [Fig. \ref{2ecti}(e-h)], since the dependence
of the charge density on $B$ is weak [Fig. \ref{2ecti}(m-p)]. On the other hand the contribution of the diamagnetic current
increases linearly to $B$ [Fig. \ref{2ecti}(i-l)].
 The energy of $L=2$ state is minimal at $B=0.26$ T. By definition
the minimal energy corresponds to zero dipole moment generated by the persistent
current. The dipole moment can be evaluated either by $dE/dB$ or
by integrating $dM={\bf j}\times {\bf r}|_z$ over the confinement plane.
For circular rings $dM$ (not shown) is simply a product of the tangential component of the current
and the distance from the center of the ring.
When the cancelation of the contributions
of the diamagnetic and paramagnetic currents to the dipole moment occurs,
the persistent current possesses two opposite current loops [see Fig.  \ref{2ecti}(c)].
The paramagnetic loop goes on the inner edge of the charge density and the diamagnetic loop at
the outer edge of the charge density. The reason for this behavior can be seen by looking at the expressions
for both the currents for circular quantum rings.\cite{odnosnikd}
Since the opposite current loops run on the edges of the persistent currents,
the magnetic fields generated by both the loops
within the charge density ring have the same orientation. In consequence the entire charge density
stays within the region of positive $B_g$ which tends to screen the external magnetic field.
Note, that in the strictly one-dimensional rings at the energy minima the paramagnetic and diamagnetic
contributions exactly cancel \cite{sigmund,szafranes} yielding a zero net current.
In the two-dimensional rings
the paramagnetic and diamagnetic contributions
cancel exactly only near $r\simeq R$ but
not on the edges of the ring. The reversal of the persistent current orientation along the radius of the circular ring
is obtained at the energy minima for all the electron numbers studied here.

\begin{figure}[ht!]
\centerline{\hbox{\epsfysize=150mm
               \epsfbox[43 157 512 728] {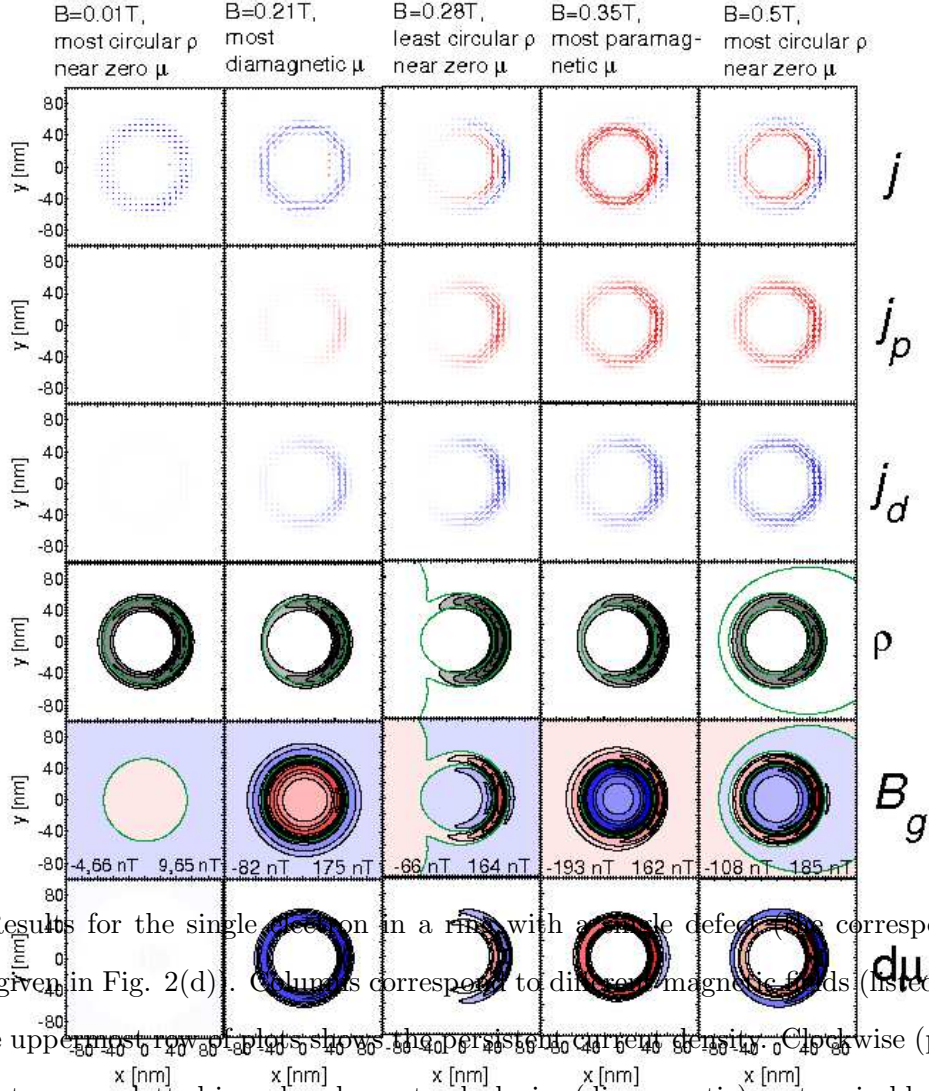}\hfill}}
\caption{Results for the single electron in a ring with a single defect (the corresponding energy spectrum given
in Fig. \ref{spectra}(d)). Columns correspond to different magnetic fields (listed on top of the plot).
The uppermost row of plots shows the persistent current density. Clockwise (paramagnetic) current vectors are plotted in red
and counterclockwise (diamagnetic) vectors in blue.
Second and third row of plots show the paramagnetic and diamagnetic contributions to the persistent current respectively.
Fourth row shows the electron density, fifth and sixth rows - the magnetic field generated by the persistent current
and local contribution to the dipole moment $dM=\left({\bf r}\times {\bf j}\right)|_z$, respectively. In both plots red (blue) region correspond to
positive (negative) $z$ components of $B$ and $dM$ vectors.
 In the plot for $B_g$ and $\rho$ the contours
corresponding to changing orientation of $B_g$ are plotted in green.} \label{m1e1z}
\end{figure}

\begin{figure}[ht!]
\centerline{\hbox{\epsfysize=150mm
               \epsfbox[21 45 575 750] {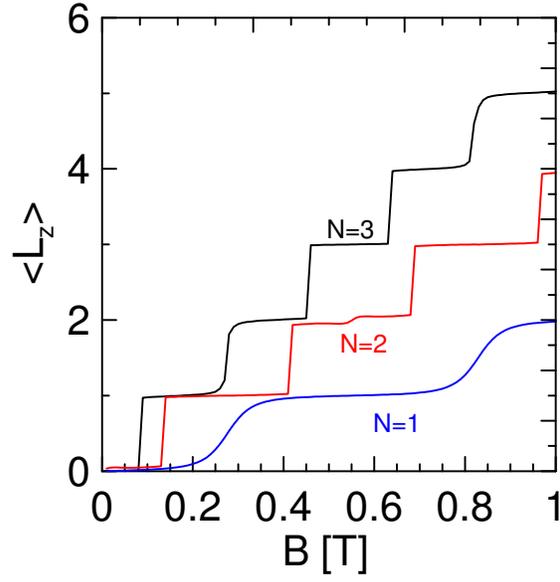}\hfill}}
\caption{Average angular momentum for the ring with a single defect in the ground state
of one, two and three electrons. }\label{l}
\end{figure}

\begin{figure}[ht!]
\centerline{\hbox{\epsfysize=80mm
               \epsfbox[208 508 550 664] {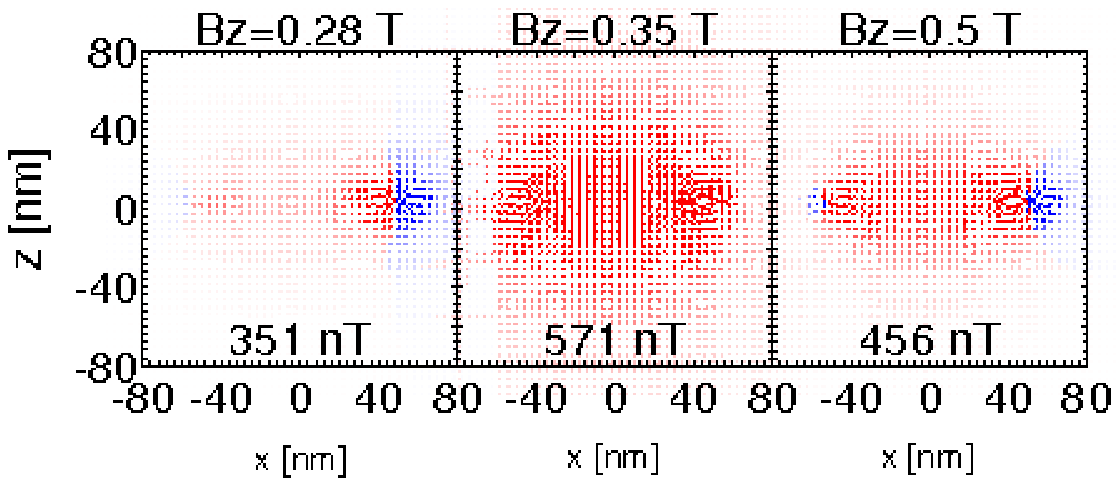}\hfill}}
\caption{
Magnetic field generated by the persistent current for a single electron ground-state in a ring with a single defect
at the $y=0$ cross section of the ring ($z=0$ is the plane of confinement). The magnetic field is plotted with red vectors in locations
in which the contribution of paramagnetic persistent currents prevails over the diamagnetic contribution.
For the dominating diamagnetic contribution we used the blue color. Maximal length of the magnetic field vector is given in the figure.
} \label{pxz}
\end{figure}

\begin{figure}[ht!]
\centerline{\hbox{\epsfysize=80mm
               \epsfbox[113 46 750 823] {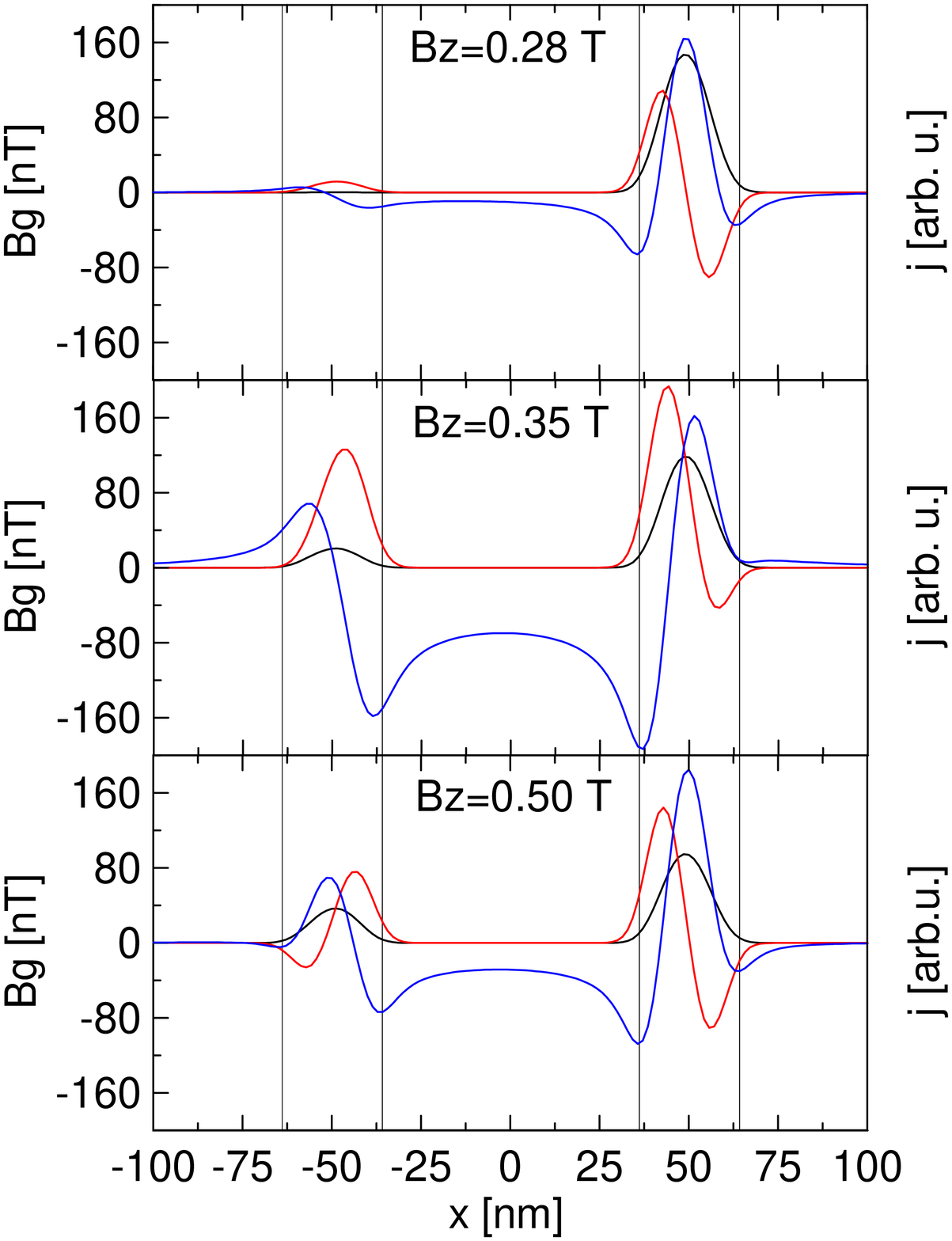}\hfill}}
\caption{
Cross section of the contour plots of Fig. \ref{m1e1z}.
Charge density (black curves), the angular component of the persistent current (red curves,
positive values corresponds to the clockwise orientation) and the $B_g$ field (blue curves)
calculated along the $x-$axis ($y=0$) for a single electron and a single defect.
} \label{cr}
\end{figure}

\begin{figure}[ht!]
\centerline{\hbox{\epsfysize=150mm
               \epsfbox[43 161 562 775] {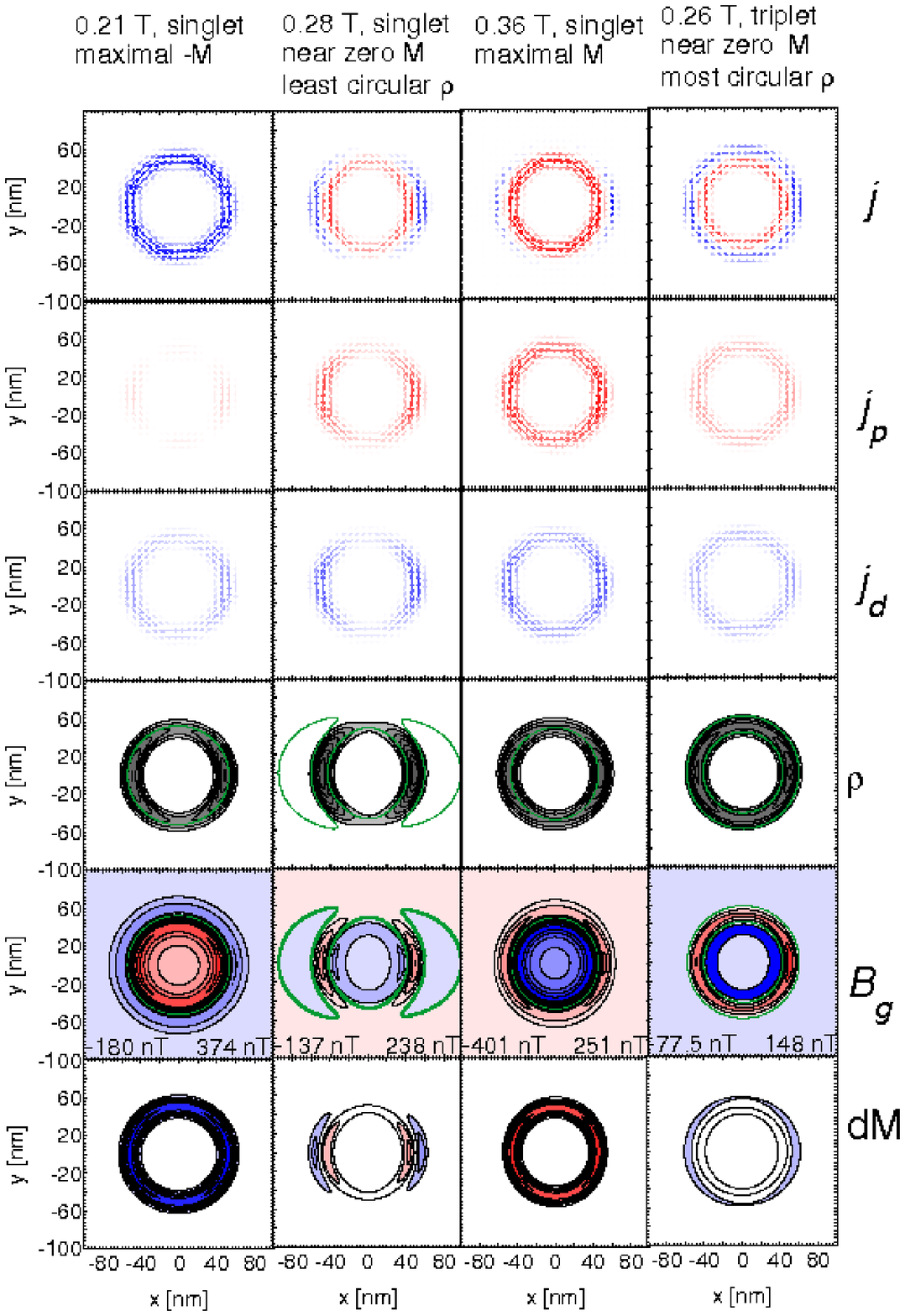}\hfill}}
\caption{Same as Fig. \ref {m1e1z} but for two electrons with the spectrum given at Fig. \ref{spectra}(e). First three columns show the results
for the lowest-energy singlet and the last column the results for the lowest-energy triplet.}\label{m1e2z}
\end{figure}

\subsection{Rings with a single defect}
Let us now consider the case when the circular symmetry of the ring is perturbed by a single attractive defect
placed in point $(x,y)=(R,0)$. The spectrum for a single electron is displayed in Fig. \ref{spectra}(d).
The crossings of energy levels of different $L$ which are observed in circular rings [Fig. \ref{spectra}(a)]
are now replaced by the avoided crossings due to mixing of angular momentum states in presence of the perturbation.
The electron density $\rho$ plot is shown in the fourth row of plots of Fig. \ref{m1e1z} (subsequent contours
from left to right correspond to increasing magnetic fields). We see that the electron density has always a maximum localized
at the defect. The strength of the electron localization near the defect oscillates with the magnetic field.
The density is most circular near the energy minima ($B=0$ and $B=0.5$ T - see. Fig. \ref{spectra}(d)]
and it deviates the strongest from circular at the energy maxima, i.e. at the center of the avoided crossings
observed for $B=0.28$ T in the spectrum.  In Fig. \ref{l} we plotted the average angular momentum calculated
in the ground-state of $N=1,2$ and 3 for the ring with a single defect. The average angular momentum
goes through plateaux of nearly integer values. At the plateaux the ground-states are nearly angular momentum
eigenstates with almost circular charge densities. The charge density deviates most strongly for the magnetic
fields for which the average angular momentum is equal to an odd multiple of $1/2$.

The dipole moment generated
by the persistent currents, which can be calculated as the derivative of the energy with respect to the magnetic field,
 vanishes for both the cases of most and least circular electron density, i.e.
both at the ground-state energy minima and the energy maxima.
In Fig. \ref{m1e1z} we additionally selected for illustration the magnetic fields for which the absolute value of
the magnetization is the largest. Second column, for $B=0.21$ T corresponds to a locally largest diamagnetic dipole moment
and the fourth column (for $B=0.35$ T) to the largest paramagnetic $\mu$.
For $B=0.01$ T we see only a residual diamagnetic current loop, and for $B=0.21$ T we notice an appearance
of the paramagnetic current. In contrast to the circular rings, both contributions ${\bf j}_p$ and ${\bf j}_d$ are
divergent, only the net current is continuous, i.e.  $\nabla \cdot ({\bf j}_p+{\bf j}_d)=0$.
Moreover, we notice a clear dependence of the paramagnetic current distribution
on the magnetic field which was to weak to be seen for the angular momentum eigenstates.
For the least circular charge density ($B=0.28$ T) the
paramagnetic current dominates in the inner edge of the electron density strongly localized at the defect
and the diamagnetic current on the outer edge, like
for circular rings.
For the least circular density the islands of positive and magnetic field density have crescent shape
and the current density goes around this island in the counterclockwise direction.
Hence, the entire charge density stays in the region of positive $B_g$ field.
We can see that the both the diamagnetic and paramagnetic current loops around the ring are broken
due to the vanishing charge density at the left side of the ring.

The field generated by the currents are plotted in the $y=0$ plane in Fig. \ref{pxz}
for the magnetic fields of vanishing dipole moment ($B=0.28$ T and $0.5$ T) and
for the maximal paramagnetic contribution ($B=0.35$ T).
The vectors were plotted with red (blue) in the regions of space for which the contribution of the paramagnetic (diamagnetic)
current is stronger.
In Fig. \ref{pxz} for $B=0.28$ T, plotted for the case when the electron localization at the defect is the strongest,
at the defect location we observe
two opposite magnetic field loops formed by the currents at the edges
of the charge density [see Fig. \ref{m1e1z}] and the magnetic field oriented parallel to the $z$ direction
in between these loops. For $B=0.5$ T we notice the magnetic field
generated by the current at the negative $x$ side of the ring (at its clean side).
For $B=0.35$ T the magnetic field loops at the ring cross section have opposite
orientations due to opposite $y$ component of the persistent current at the section. The magnetic field generated by the
electron confined within the ring at $B_z=0.35$ T is
strong and parallel to the external field within the current loop. Outside the loop $B_g$ is opposite
to the external field but weak.

The ground-state electron density deviates most strongly of the circular symmetry when
maximal mixing of subsequent $L$ states [cf. Fig. \ref{l}] occurs. Outside the center of
the avoided crossings, the profile of $B_g$ distribution resembles the one
obtained for circular ring. In fact with the exception of $B=0.28$ T, the other magnetic fields
selected for illustration in Fig. \ref{m1e1z} correspond to ground states which are nearly angular momentum eigenstates (see Fig. \ref{l}).
In all the plots presented in Fig. \ref{m1e1z}, {\it most}
of the electron density stays within positive $B_g$ region, in which the net current tends to screen the external field.
In Fig. \ref{cr} we plotted a cross section of the contour plot of Fig. \ref{m1e1z}
along the $y=0$ axis. For the least circular density at $B=0.28$ T we see that the maximum
of the density coincides with the maximum  $B_g$, although the edges of the charge density
stay in the region of  $B_g$ parallel to the external field. Both $B=0.28$ T and $B=0.5$ T
correspond to vanishing dipole moment. In the first case the persistent current loop around
the ring is broken and
 the generated external field is mainly localized at the defect.
For the other $B$ value, for the most circular charge density, non-zero values of $B_g$ are found inside the ring,
since both the diamagnetic and paramagnetic current circulate around it.
In all the presented cases we can see that the leakage of the magnetic field $B_g$ outside the ring is quite weak.
The last row of Fig. \ref{m1e1z} shows the local contribution to the dipole moment of the ring.
Zero dipole moment (obtained for $B=0$, near $B=0.28$ T and near $B=0.5$ T) is produced when the positive and negative fluxes
of $dM$ are equal.

In circular rings containing two electrons the magnetic field induces transitions of the ground-state
spin and angular momentum. For the single defect considered here the angular momentum and the spatial parity
are not good quantum numbers, but the spin still is.  One observes the ground-state spin transitions [Fig. \ref{spectra}(e)]
in function of the magnetic field, like in the circular ring case.
For a given spin value we observe
avoided crossings of energy levels instead of crossings.
For instance, in case of a circular ring there is a crossing of the lowest-energy singlets $L=0$ and $L=2$ near
$B=0.28$ T [see Fig. \ref{spectra}(b)]. For the ring containing the defect, an avoided crossing
is opened replacing the angular momentum transition [see Fig. \ref{spectra}(c)].
Fig. \ref{m1e2z} shows the densities, currents and magnetic properties for the two-electron ring with a single defect.
 Although the potential of a single defect breaks the
parity symmetry of the confinement potential,  we obtain charge densities which are nearly symmetric
with respect to the point inversion through the origin. One of the electrons
becomes pinned at the defect and the other stays on the other side of the ring
repulsed by the potential of the pinned electron.
The extent of the localization of both electrons pulsates
with the magnetic field like for the single electron and is the strongest when the energy
of the state for a given spin value is maximal.
The difference to the single-electron problem is that the densities which
correspond to the maximal deviation off the circular symmetry do not correspond to the ground-state.
In this case
for the lowest-energy singlet with the least circular density the ground-state corresponds to the
spin triplet with the most circular density and vice versa.
However, for higher magnetic fields the Zeeman effect removes the singlets from the ground state (see below),
so the triplets with the locally maximal energy and strongest localized electrons appear in the ground-state.

\begin{figure}[ht!]
\centerline{\hbox{\epsfysize=150mm
               \epsfbox[88 69 408 750] {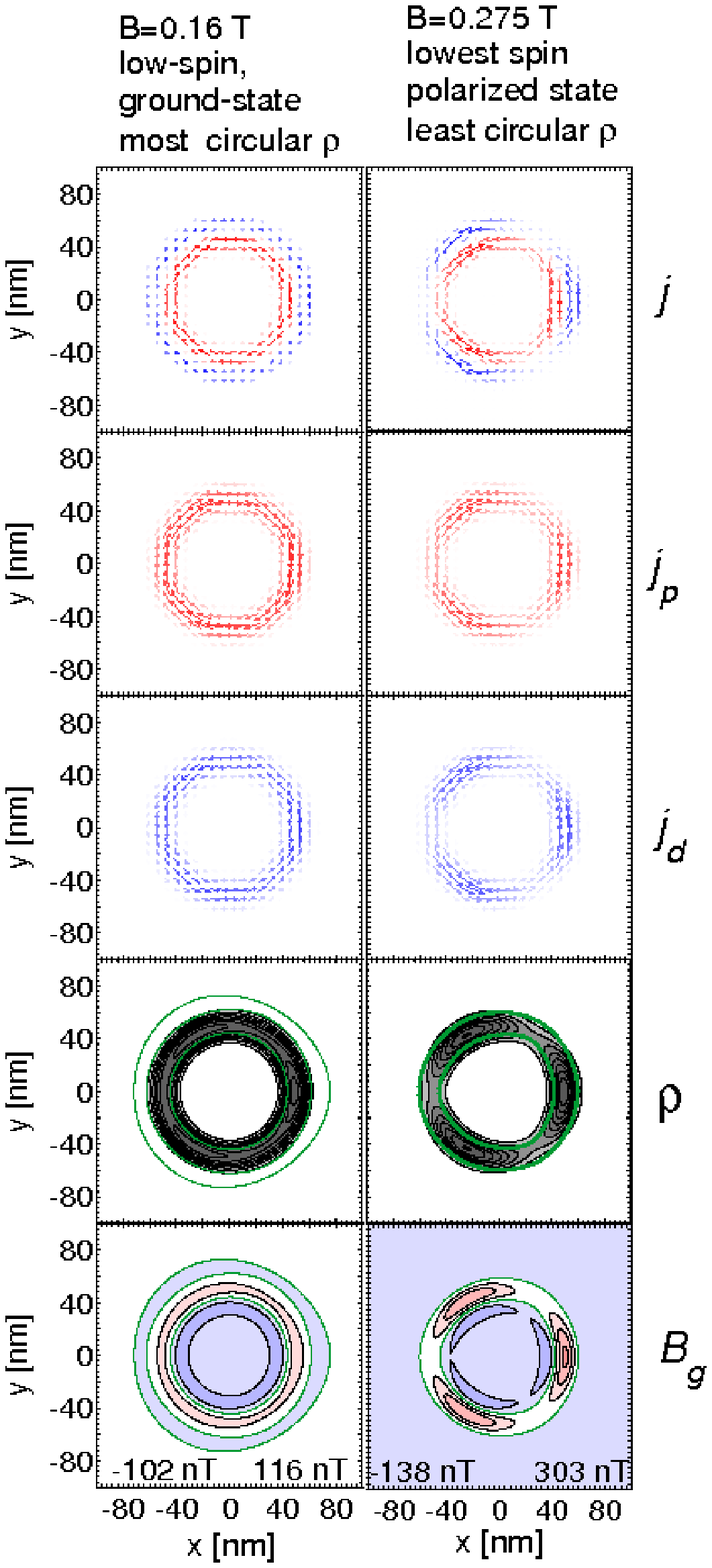}\hfill}}
\caption{Same as Fig. \ref {m1e1z} but for three electrons with the spectrum given at Fig. \ref{spectra}(f).
The left column corresponds to the low-spin ground state and the magnetic field for which the energy is minimal
(dipole moment is zero). The right column corresponds to the lowest-energy spin-polarized state for the magnetic field
corresponding to the center of an avoided crossing of the two lowest-energy spin-polarized states.
}\label{m1e3z}
\end{figure}

Fig. \ref{m1e3z} shows the results for the cases of balanced diamagnetic and paramagnetic
currents. The left column of Fig. \ref{m1e3z} corresponds to the minimal energy [Fig. \ref{spectra}(f)] of the low-spin ground state
(most circular charge density),
and the right column to the lowest-energy spin-polarized state when its energy is maximal (most localized charge density) as a function of the magnetic field.
In the left column the density possesses three maxima with one localized at the center of the defect, but the maxima
are not very pronounced. In fact, the density is nearly ideally circular. Comparing this density with the
''most circular'' densities obtained for $N=1$ and $N=2$ we conclude that the density
obtained at the minimal ground-state energy becomes more ideally circular with the growing electron number.
This can be understood as due to screening of the attractive defect potential by the confined charge.
In the single-electron picture the defect potential for the second electron and third electron added to the ring is partially screened by the Coulomb potential
of the electron more or less strongly localized at the defect.
Non-circular electron density is associated to non-integer values of the average angular momentum  [Fig. \ref{l}].
We can see that for the two and three electrons the magnetic field range between the integer-valued plateaux
is significantly reduced with respect to $N=1$,
which is also related to the screening of the defect.

In Fig. \ref{l} we notice that for three electrons the first step from $\langle L_z\rangle=0$ to 1 is abrupt
and the next one from $\langle L_z\rangle=1$ to 2 is smooth. The abrupt step is due to the spin ground-state transition
at $N=3$. The second step is smooth since the two low-spin energy levels that appear to cross near $B=0.28$ T in the energy
spectrum [Fig. \ref{spectra}(f)] in fact enter into very narrow avoided crossing of width of the order of $\mu$eV.
The continuous character of the ground-state energy level at these avoided crossings is visible in the plot
of the average angular momentum in Fig. \ref{l} (see also below for the plot for the derivative of the ground-state
energy with respect to the magnetic field in Fig. \ref{dip}).
A small width of these avoided crossing is remarkable when compared to the wide avoided crossings that appear between
the ground-state branch of energy levels and the excited part of the spectrum near $E=-578.4$ meV.
The difference in the width of the crossing can be understood if we remind the fact that the
charge density of each state is the most circular near the minimum of the energy [compare
for instance the densities obtained for $B=0$ (energy minimum) and $B=0.21$ T (largest $dE/dB$)
in Fig. \ref{m1e2z}]. The states that nearly
cross in the ground-state energy branch are therefore almost  eigenstates of the angular momentum, hence
the small width of the avoided crossing. Outside the energy minima the densities are more susceptible to the
confinement potential deformations. The avoided crossings for the $N=3$ low-spin ground-states
are much tighter than for the single electron [Fig. \ref{spectra}(d)] which is again related to the screening
of the defect by the confined electrons.

Let us now turn our attention to the currents and the generated fields for the three electron system in a ring with a single defect.
For the ground-state (left column of Fig. \ref{m1e3z}) at the energy minimum there are two opposite current loops going around the ring at the edges
of the structure. For the Wigner crystalized charge density (plot for the triplet at $B=0.275$ T in Fig. \ref{m1e3z})
we notice that the net current loop around the ring is nearly broken in places when the electron density is minimal.
Instead we observe three current vortices around each single-electron charge density island.
Within each of the vortices the current circulates counterclockwise producing the
$B_g$ which is opposite to the external field within each density island.

\subsection{Rings with two symmetrically placed defects}

\begin{figure}[ht!]
\centerline{\hbox{\epsfysize=150mm
               \epsfbox[21 45 575 750] {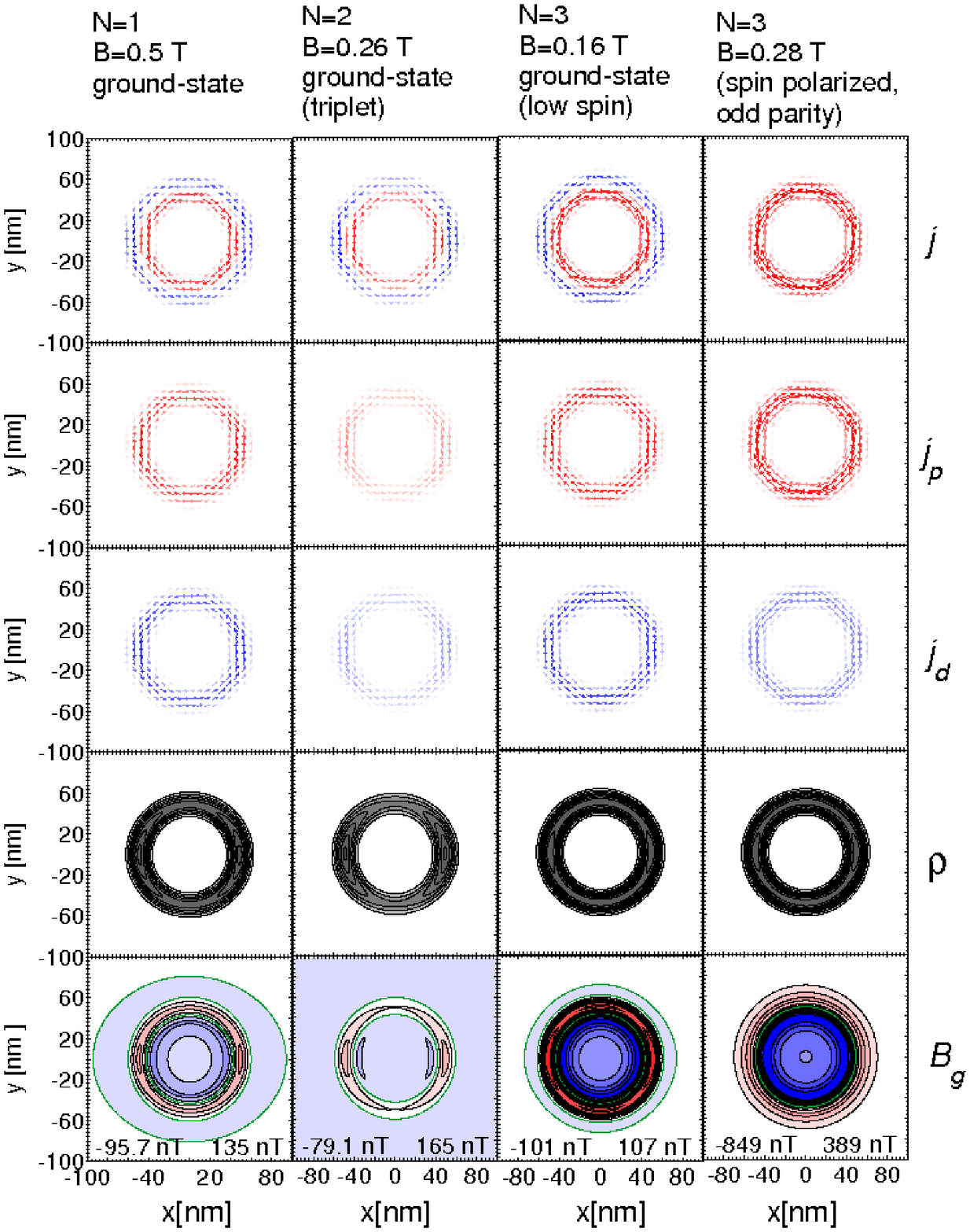}\hfill}}
\caption{Net persistent current ($j$), the paramagnetic ($j_p$) and diamagnetic ($j_d$)
contribution, charge density $\rho$ and  $B_g$ -- the magnetic field generated by the persistent current
for a ring with two defects placed symmetrically in $(x,y)=(R,0)$ and $(-R,0)$ points.
The columns correspond to various number of electrons $N$ and various magnetic fields $B$ which are listed
at the top of each column. The first three columns at left correspond to ground states and
nearly vanishing magnetization, the fourth column
corresponds to an excited state.
The first column can be compared with the results presented on the last column in Fig. \ref{m1e1z} for a single defect
(same $B$).
The results for a single defect corresponding to $N$ and $B$ of the second column are presented in the last
column of Fig. \ref{m1e2z}.  The last two columns (for three electrons) can be compared with Fig. \ref{m1e3z} for a single defect.
}\label{rdtdd}
\end{figure}

\begin{figure}[ht!]
\centerline{\hbox{\epsfysize=150mm
               \epsfbox[39 63 491 754] {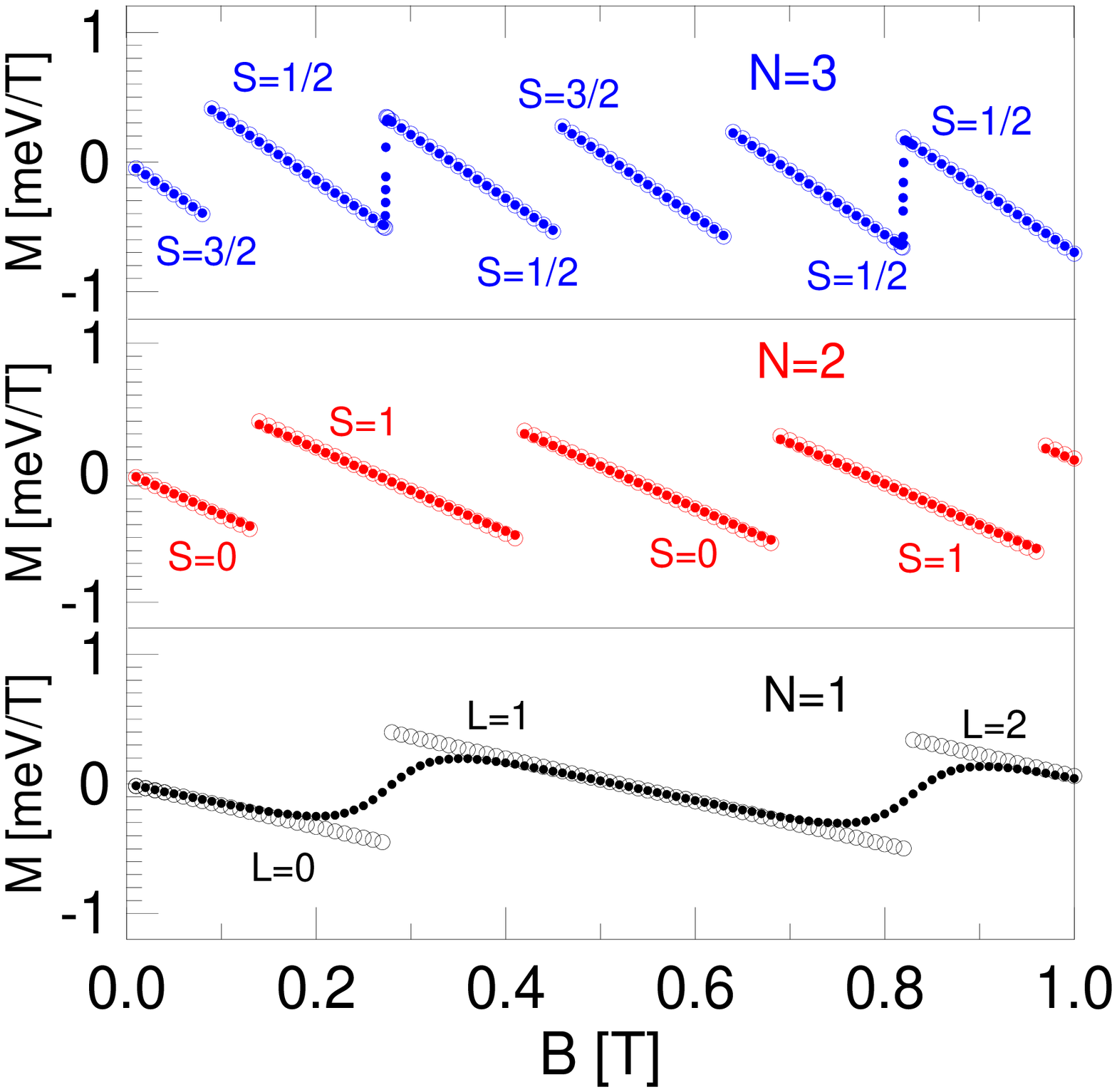}\hfill}}
\caption{Dipole moment as function of the magnetic field
for $N=1$, $N=2$ and $N=3$ electrons. Full dots show the results for a single
defect and open dots for the clean quantum ring.
In plot for $N=1$ we list the ground-state angular momentum for the clean ring $L$.
In plots for $N=2$ and 3 we give the z-component of the total spin ($S$).} \label{dip}
\end{figure}

We consider next the ring in which a second defect is introduced at a position placed symmetrically with
respect to the first one as indicated in Fig. \ref{pot}. Introduction of the second defect restores the parity
symmetry of the confinement potential which was lifted by the first defect. The system with two defects placed symmetrically
agrees in symmetry with the elliptical quantum rings.\cite{foomin}
The consequences of the symmetry restoration for the energy spectra can be seen by comparison of Fig. \ref{spectra}(d-f) with
Fig. \ref{spectra}(g-i).  For the single electron in the ground-state we find crossings of energy levels
of even and odd parity [Fig. \ref{spectra}(g)] in contrast to the avoided crossings observed for a single defect [Fig. \ref{spectra}(d)].
At the crossing of energy levels the dipole moment generated by the ground-state single-electron current changes abruptly
from diamagnetic to paramagnetic. For a single defect the transition was smooth with the electron maximally localized at
the defect for the magnetic field corresponding to vanishing dipole moment -- balanced paramagnetic and diamagnetic currents.
The currents, charge density and the magnetic field for $N=1$ at $B=0.5$ T are presented in
the first column of Fig. \ref{rdtdd}. This value of $B$ corresponds to the local energy minimum and
vanishing magnetization. The plots can be compared with the results obtained for a single defect
and presented in the last column of Fig. \ref{m1e1z}.

For $N=2$ the consequences of the parity restoration are not very pronounced. The ground-state spin oscillations
are for two-defects accompanied by the spatial symmetry transitions [Fig. \ref{spectra}(h)]. For a single
defect the currents, electron density and generated magnetic fields were nearly symmetric with respect
to inversion ${\bf r}\rightarrow -{\bf r}$. Now, the symmetry is perfect -- compare the last column of Fig. \ref{m1e2z}
with the second column of Fig. \ref{rdtdd}.

For three electrons restoration of the parity symmetry transforms the very narrow avoided crossings in the
low-spin ground-states (see the discussion above) into crossings of odd and even energy levels.  Moreover the avoided crossings between states
separating the ground-state energy level branch from the excited part of the spectrum characteristic
to the single defect [Fig. \ref{spectra}(f)], are closed when the symmetry of the confinement potential
is restored [Fig. \ref{spectra}(i)]. For a single-defect
the lowest-energy spin-polarized ground-state correspond to a pinned Wigner molecule (like the one observed in the
right column of Fig. \ref{m1e3z}) with a single
electron localized at the defect forming an equilateral triangle with the most probable positions of two other electrons.
For two defects the Wigner crystallization of the charge density in the laboratory frame is not observed.
The classical Wigner molecule has two equivalent charge distributions with one of the electrons
localize in the left or right defect. None of these classical configurations can appear in the laboratory frame
since it would result in breaking the symmetry of the confinement potential by the charge density.
Maximal electron localization for the pinned Wigner molecule occurs at the avoided crossing of the spin-polarized energy levels.
When the parity is restored these avoided crossings are no longer observed. Instead, they are replaced by crossings of energy
levels since the subsequent spin-polarized ground states correspond to opposite parities.
Results for the charge density and currents for three electrons in a ring with a single defect [Fig. \ref{m1e3z}] should
be compared to the results for two defect in the last two columns of Fig. \ref{rdtdd}. For two defects the charge density is nearly
circular in both presented magnetic fields.

\begin{figure}[ht!]
\centerline{\hbox{\epsfysize=45mm
               \epsfbox[39 250 458 610] {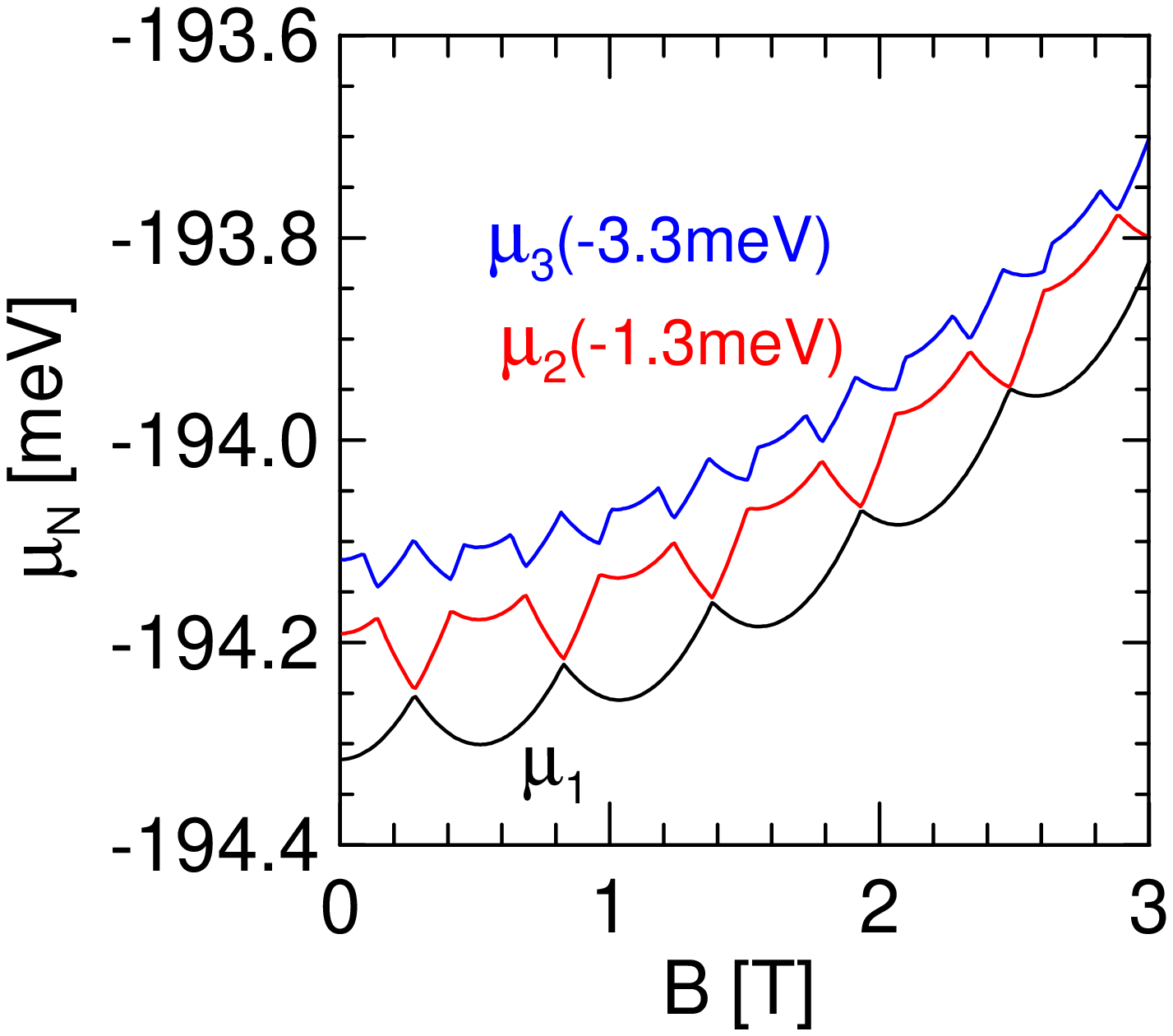}a)\epsfysize=45mm
               \epsfbox[39 250 458 610] {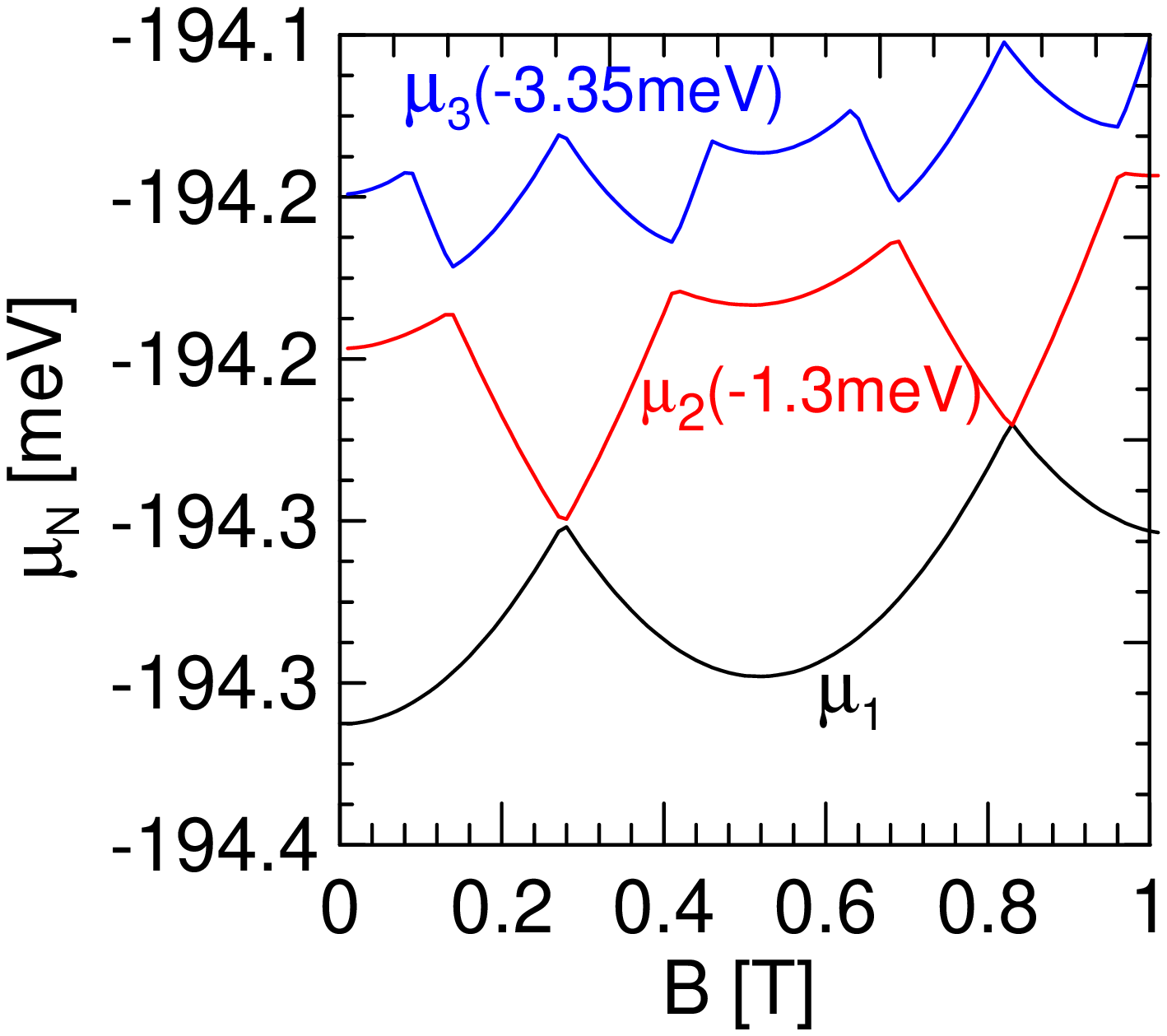}b)\epsfysize=45mm
               \epsfbox[39 250 458 610] {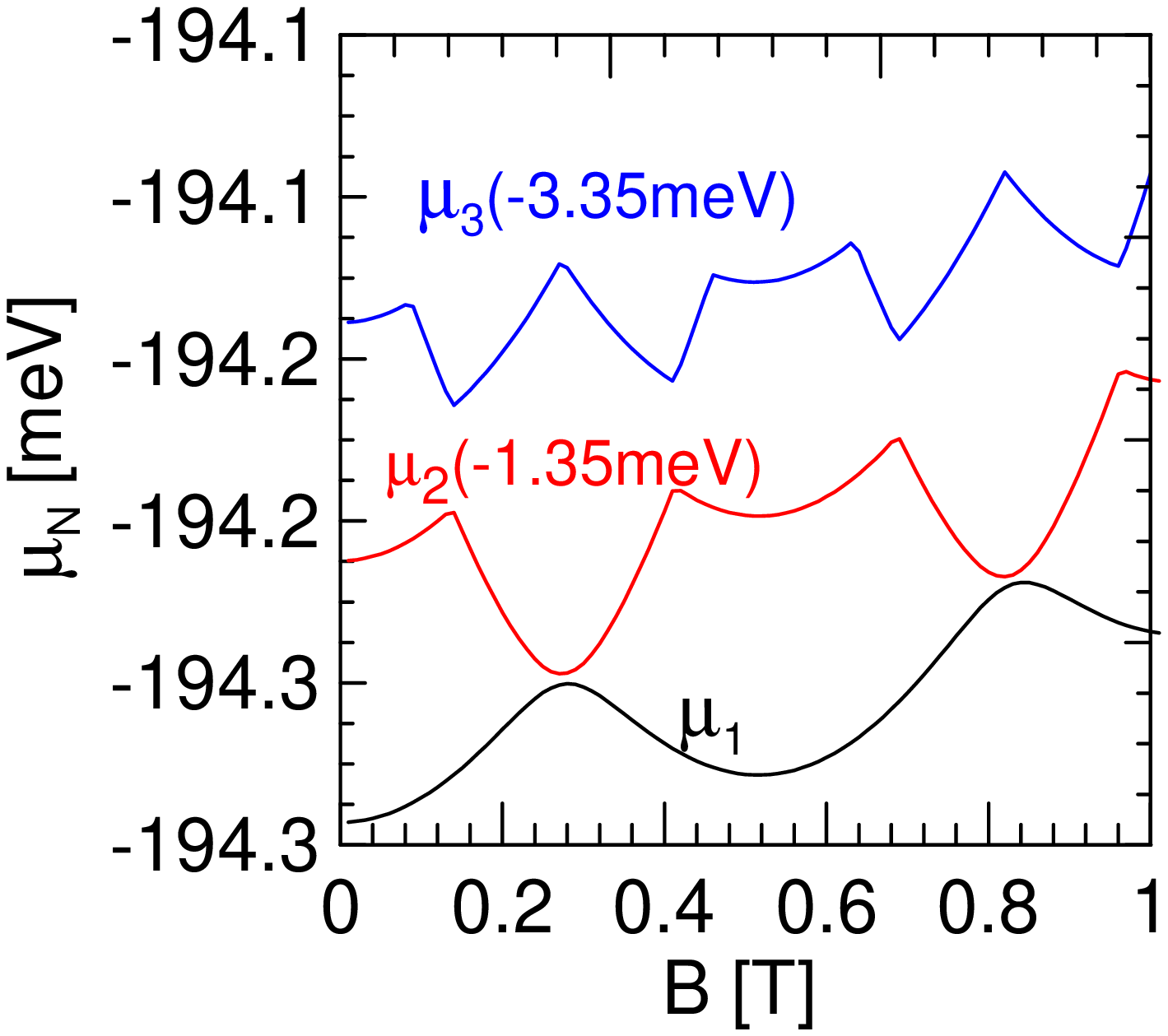}c)
               \hfill}   \vspace{0.5cm}}
\caption{Chemical potential for one, two and three electrons in a clean ring (a), in a ring with a single
defect (b) and with two defects (c). Plots for $\mu_2$ and $\mu_3$ were for clarity shifted on the
energy scale by the values given in the plots.}
\label{cp}
\end{figure}

In Fig. \ref{dip} we plotted the dipole moment as a function of the magnetic field
for one, two and three electrons in a clean ring (open dots) and the ring with a single
defect (full dots). For the clean ring we observe discontinuities at the ground-state angular
momentum transitions. For each ground-state angular momentum the dipole moment acquires
diamagnetic character when $B$ increases. For the ring with the defect the dipole moment
for a single electron becomes continuous as the avoided crossings are opened in the spectrum.
For two and three electrons the results for the clean ring and the ring with the defect are nearly identical.
Differences are only visible for three electrons: when $\mu$ becomes a continuous function of $B$
at the narrow avoided crossing that appear in the low-spin ground state.

Chemical potentials -- defined as the difference
in the energies of the confined system $\mu_N=E_N-E_{N-1}$ --  are displayed for the clean ring in Fig. \ref{cp}(a) and for rings with a single in Fig. \ref{cp}(b)
or two defects in Fig. \ref{cp}(c).  For the clean ring [Fig. \ref{cp}(a)] the chemical potential of the $N$ electron system has $\Lambda$ shaped cusps
at the angular momentum / spin transitions. The $\Lambda$ cusp for $N$ electrons is translated into a V cusp on the chemical potential
of $N+1$ electrons. Both types of cusp form a characteristic pattern due to the fractional Aharonov-Bohm effect.
The chemical potential for two electrons exhibits two $\Lambda$ cusps for a single $V$ cusp, and $\mu_3$ has
three $\Lambda$ cusps for two V cusps.
For the ring with a single defect [Fig. \ref{cp}(b)] the chemical potential of the first electron becomes
smooth, and the V cusps disappear of the $\mu_2$ plot replaced by a U shaped minimum.
The maximum of the chemical potential for three electrons near $B=0.28$ T
is smooth, although it is hardly visible at the scale of the plot.
The smooth maximum results of the narrow avoided crossing of the low-spin $N=3$ ground state.
When the parity symmetry is restored by addition of the second defect [Fig. \ref{cp}(c)], the chemical potential patters acquires
the same character as for the circular ring.

\subsection{High magnetic fields and the spin Zeeman effect}

\begin{figure}[ht!]
\centerline{\hbox{\epsfysize=45mm
               \epsfbox[39 250 458 610] {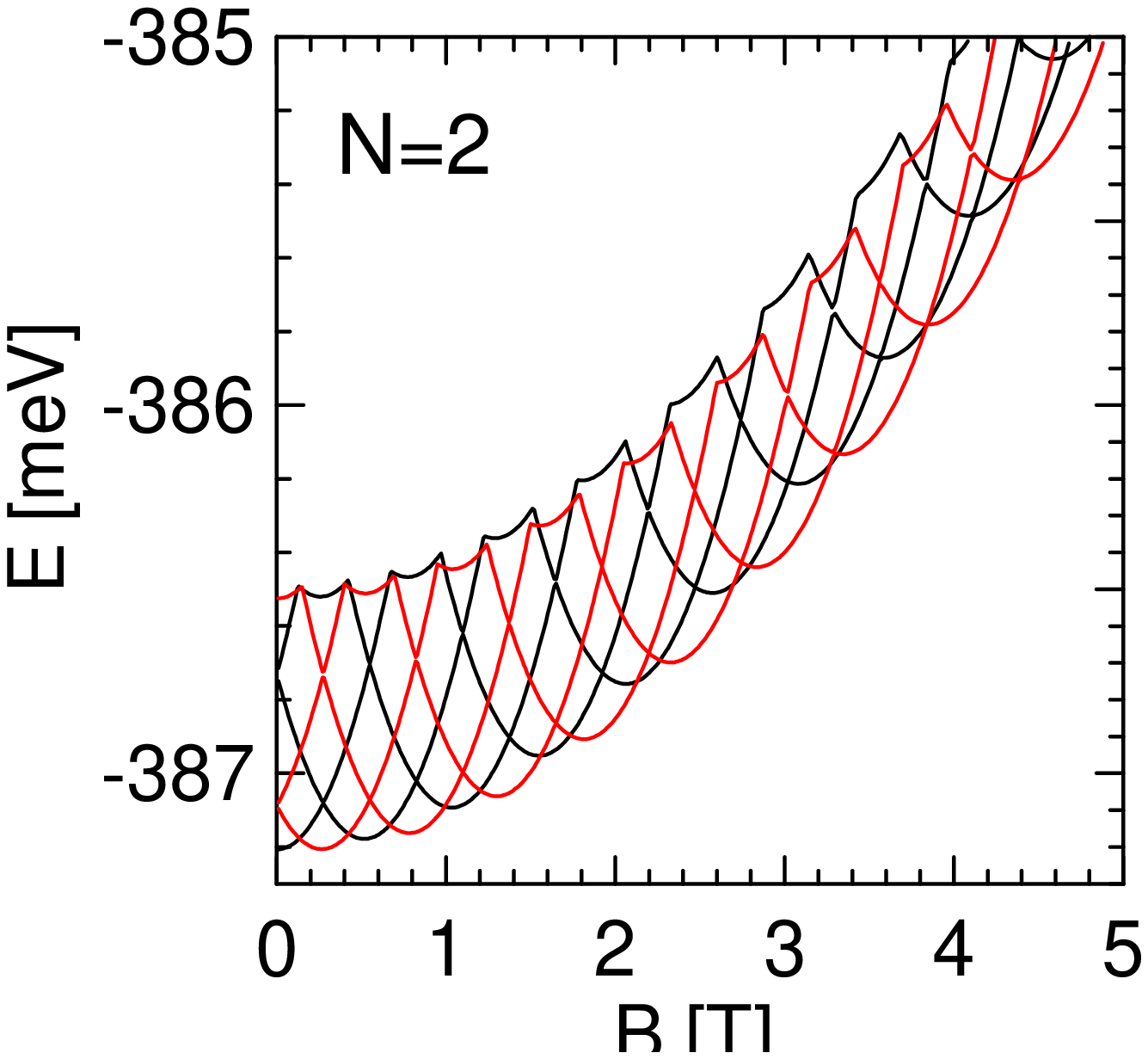}a)\epsfysize=45mm
               \epsfbox[39 250 458 610] {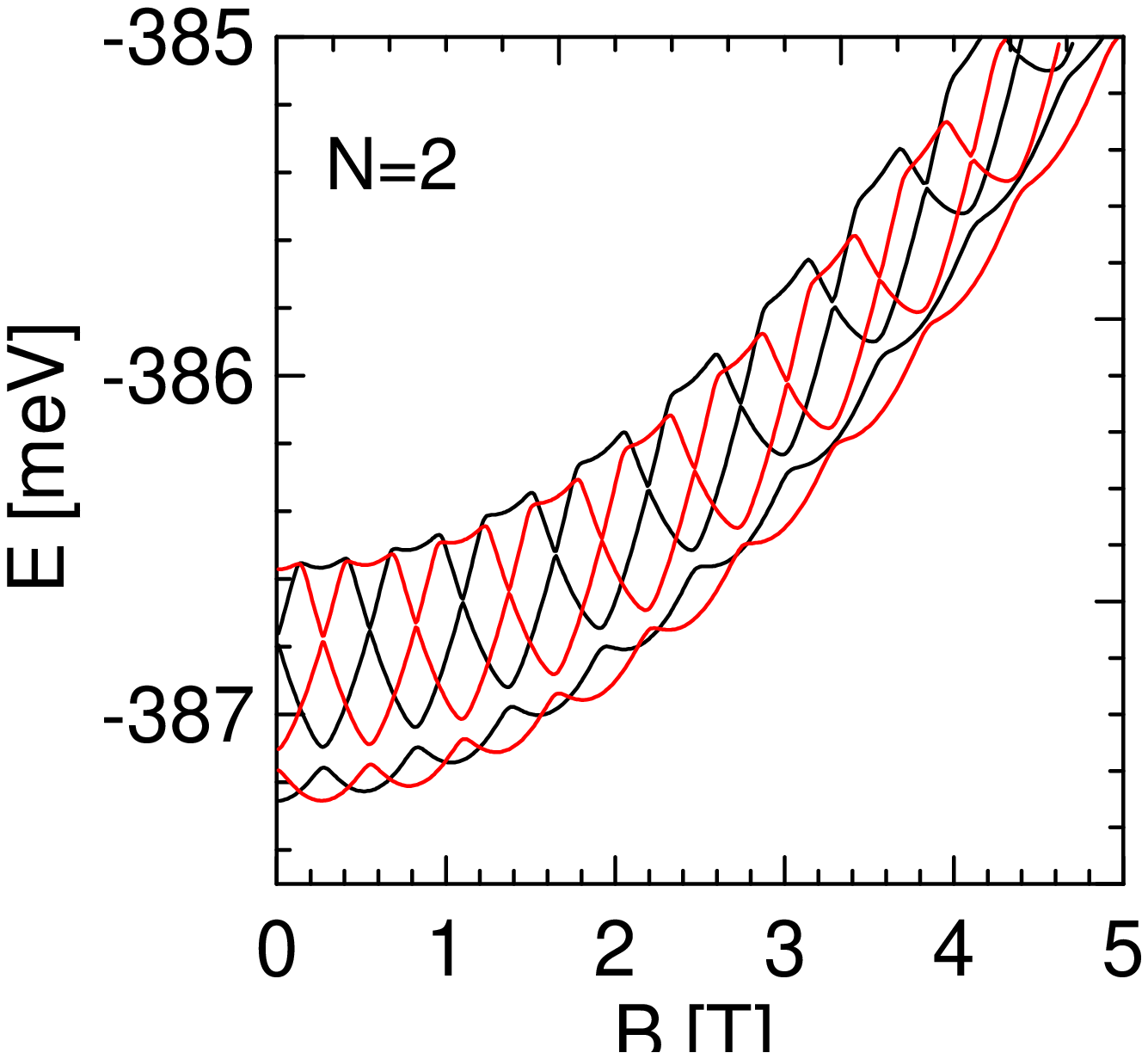}b)\epsfysize=45mm
               \epsfbox[39 250 458 610] {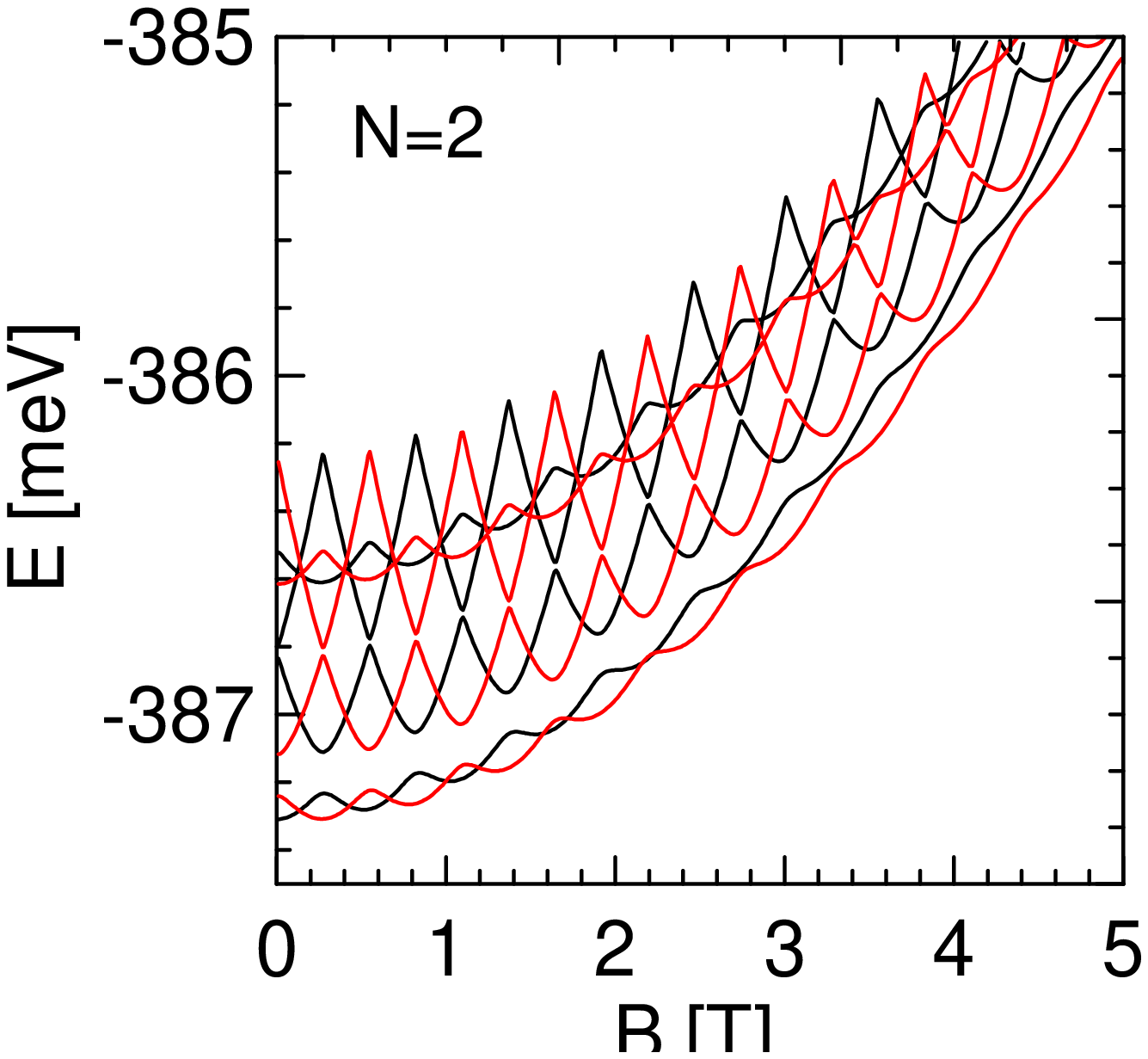}c)
               \hfill}   \vspace{1cm}
               }
               \centerline{\hbox{\epsfysize=45mm
               \epsfbox[39 250 458 610] {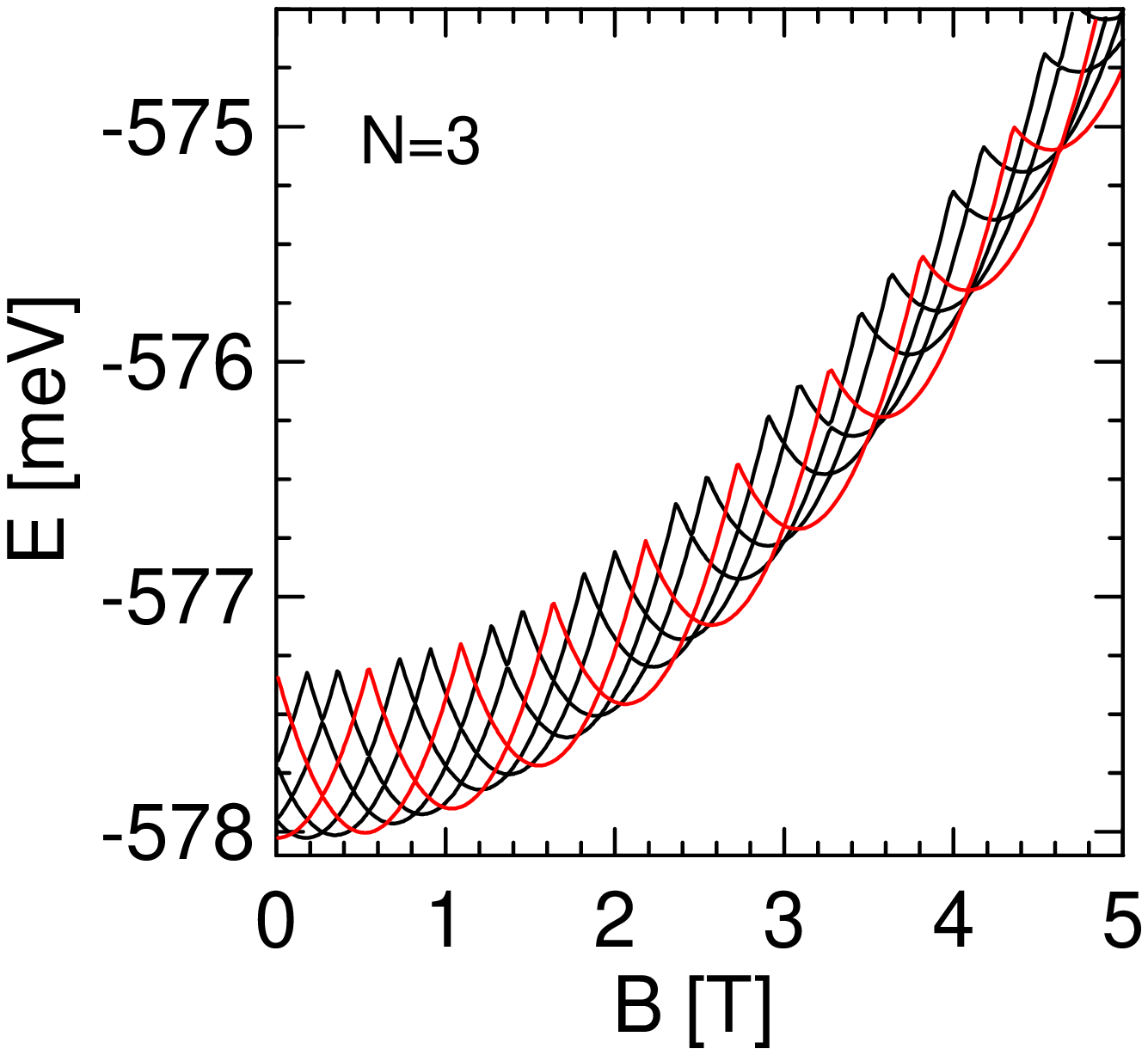}d)\epsfysize=45mm
               \epsfbox[39 250 458 610] {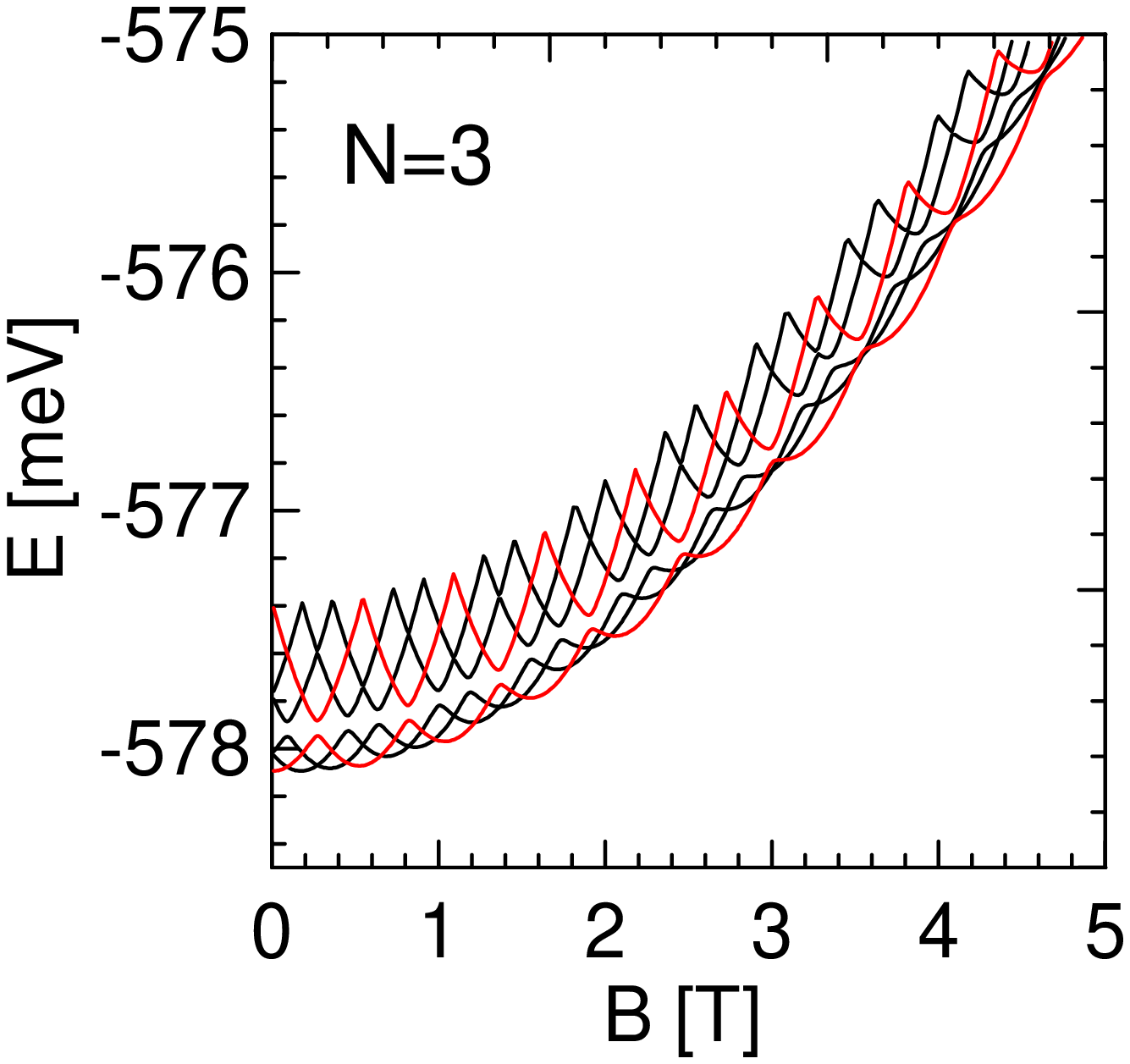}e)\epsfysize=45mm
               \epsfbox[39 250 458 610] {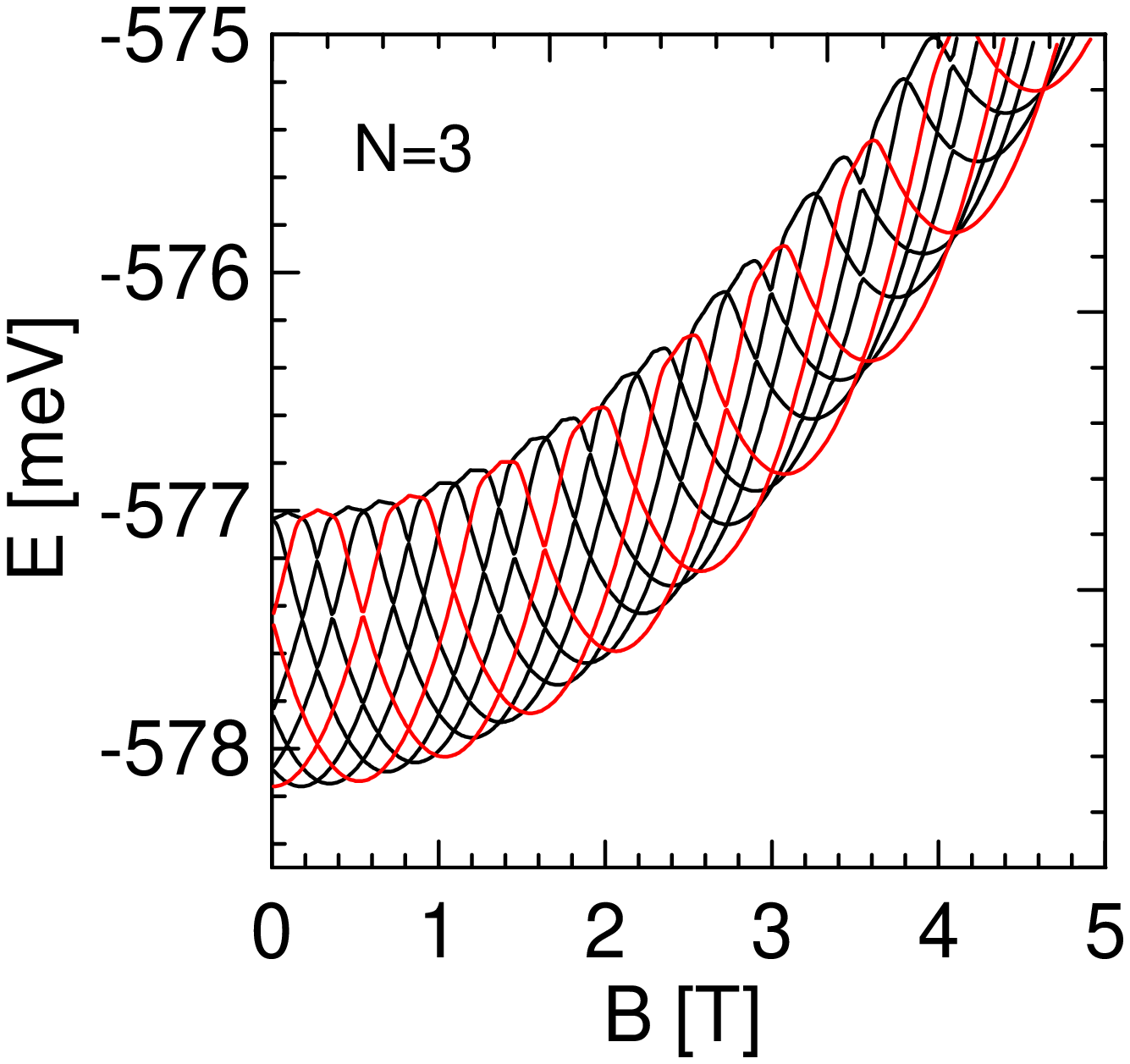}f)
               \hfill}
               }
\caption{Energy spectra for two (a,b,c) and three (d,e,f) electrons
with the spin Zeeman effect included for the clean ring (a,d), the ring with a single (b,e) and two (c,f)
defects. For $N=2$ we plot with the red lines the triplet energy levels corresponding to $S_z=-1$ (triplets with $S_z=0$ and $S_z=1$ correspond
to higher energies). For $N=3$
the black energy levels correspond to $S=1/2$ and $S_z=-1/2$ and the red energy levels to $S=3/2$ and $S_z=-3/2$.}
\label{spectraz}
\end{figure}

Fig. \ref{spectraz} shows the spectra calculated for higher magnetic fields
with the spin Zeeman effect,  or $N=2$ (a-c) and $N=3$ (d-f) electrons.
The first column of the plots [Fig. \ref{spectraz}(a,d)] corresponds
to a clean quantum ring, the middle column  [Fig. \ref{spectraz}(b,e)] to the ring with a single
defect
and the last one to the ring with two defects [Fig. \ref{spectraz}(c,f)].
In Fig. \ref{spectraz} we plotted with the red color the spin polarized
energy levels with the most negative $S_z$ component ($S_z$=-1 for two electrons
and $S_z=-3/2$ for three electrons). Spin polarized states of other $S_z$ values are always
higher in the energy. We see that spin-polarized states becomes the ground-state at higher
magnetic field, and the regularity of the ground-state spatial and spin symmetry transformations known
from Fig. \ref{spectra} is perturbed by the Zeeman effect, which favors the spin-polarized state.
At higher magnetic fields the ground state becomes permanently spin polarized.

\begin{figure}[ht!]
\centerline{\hbox{\epsfysize=45mm
               \epsfbox[39 250 458 610] {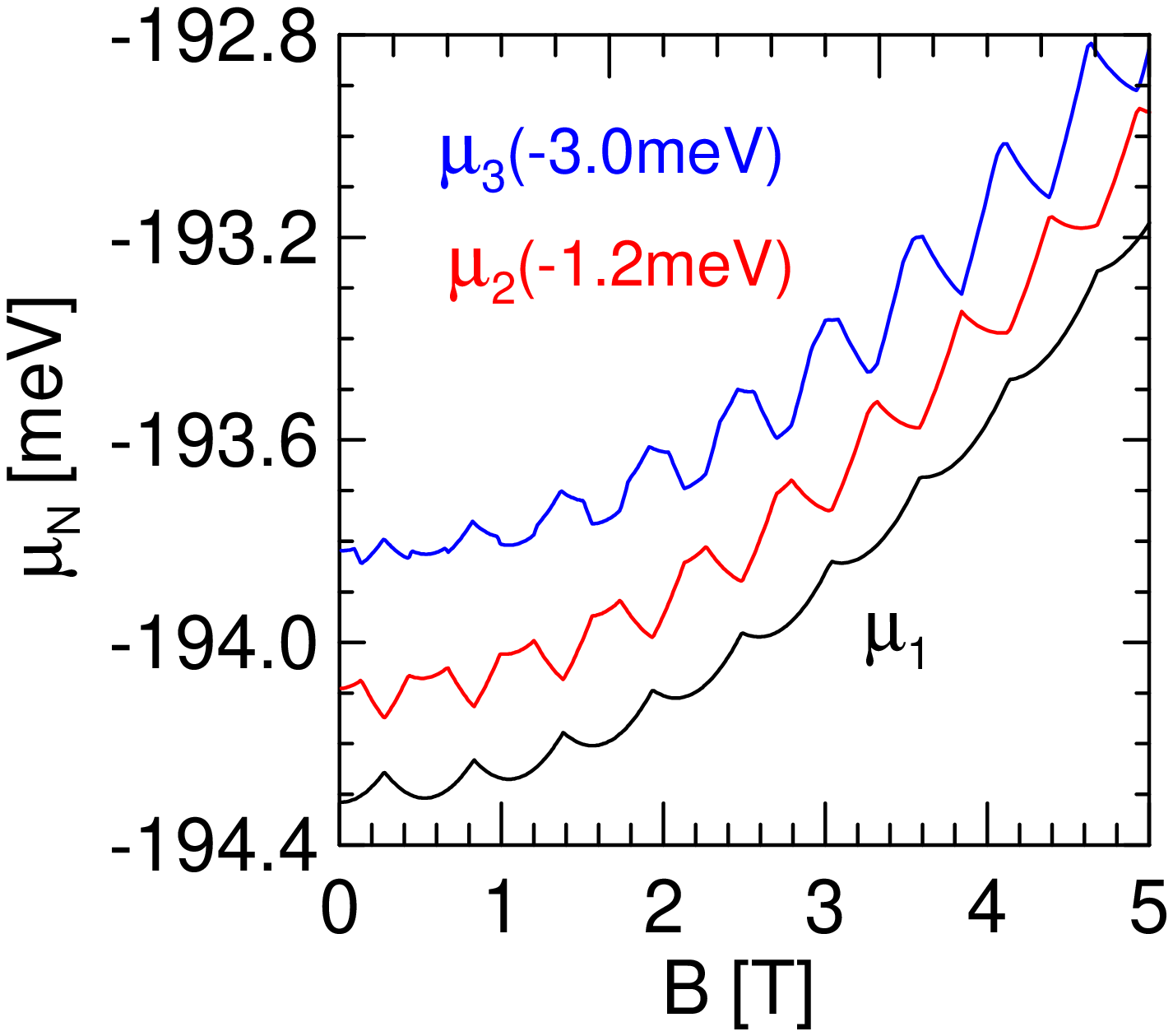}a)\epsfysize=45mm
               \epsfbox[39 250 458 610] {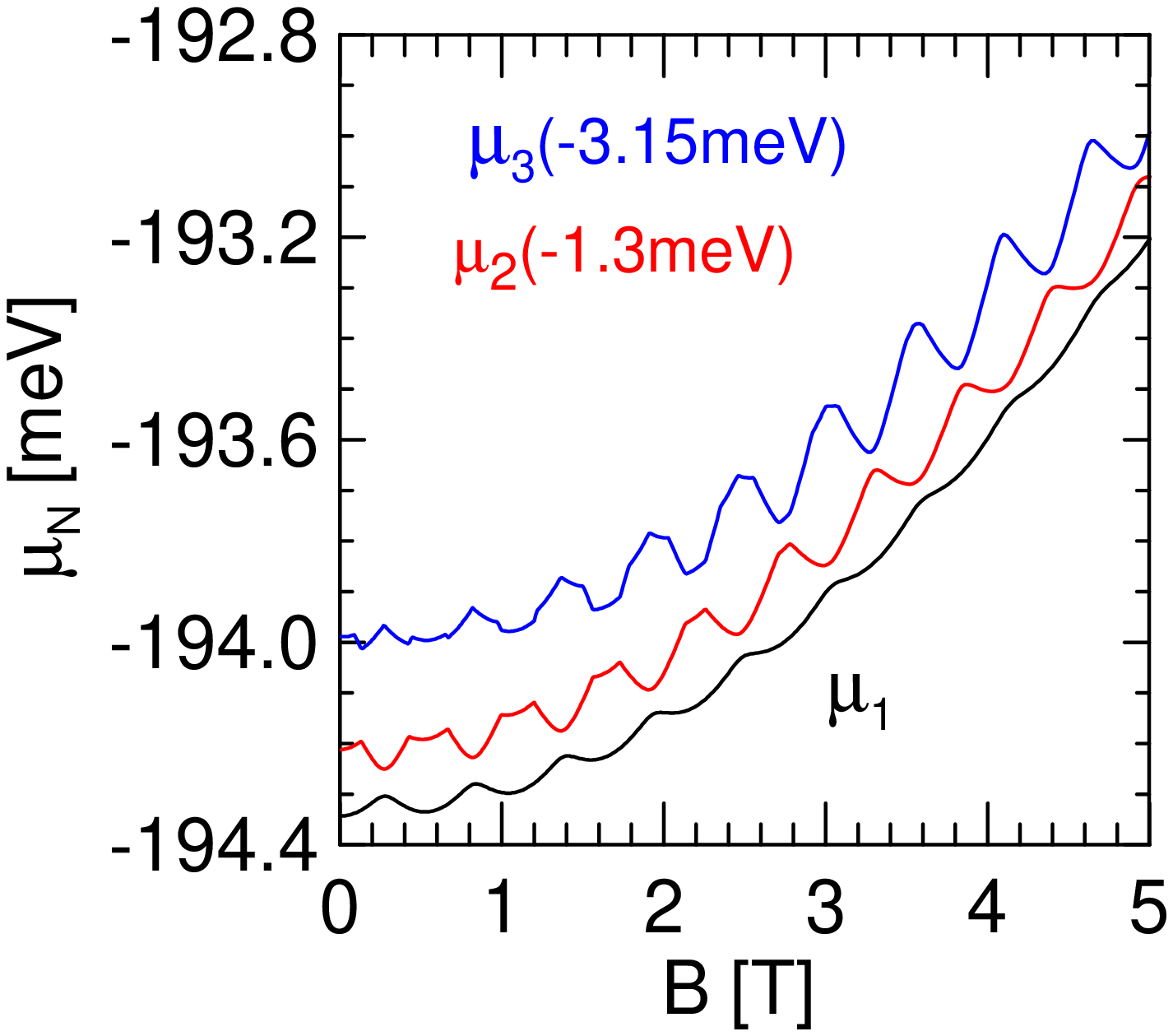}b)\epsfysize=45mm
               \epsfbox[39 250 458 610] {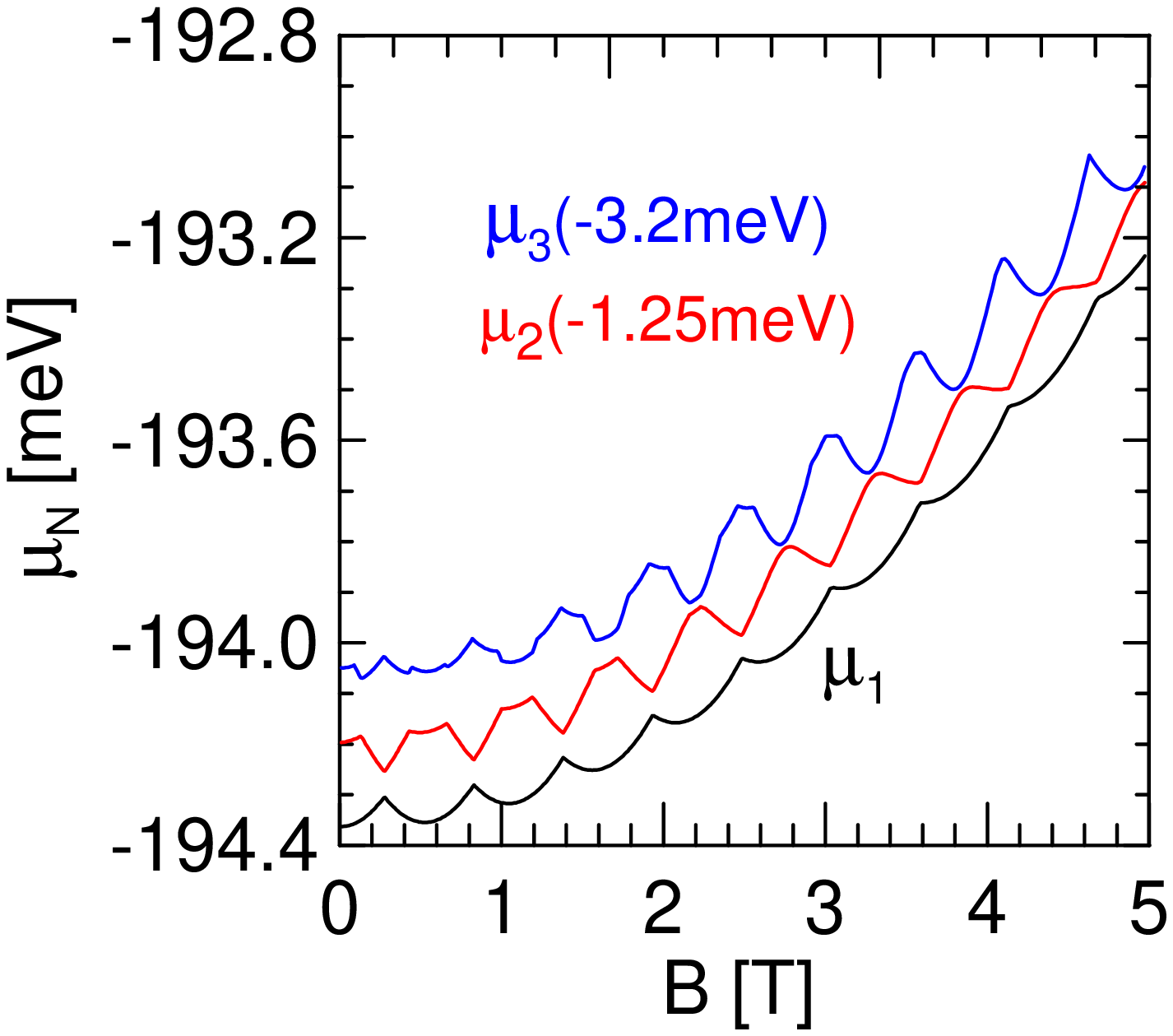}c)
               \hfill}   \vspace{0.5cm}
               }
\caption{Chemical potentials for a clean quantum ring (a), for the ring with a single defect (b) and for
the ring with two deffects (c) with the account taken for the Zeeman effect.}
\label{pcz}
\end{figure}

The plots of chemical potentials are given in Fig. \ref{pcz}. For all the considered cases of the ideally circular ring [Fig. \ref{pcz}(a)]
and the rings with defects [Fig. \ref{pcz}(b,c)] the structure of the cusps at high fields becomes relatively simple.
For the circular ring at high field for both $N=2$ and $N=3$ we observe a single $\Lambda$ cusp for a single $V$ cusp.
In the absence of the Zeeman effect the ground-state angular momentum increases by one at each ground-state transformation.
The spin Zeeman effect at high magnetic field produces the ground states of odd $L$ for $N=2$ and multiples of 3 for $N=3$.
This results in removal of the fractional character of the Aharonov-Bohm oscillation at high magnetic field.
We observe that the ground-state transformations for $N=2$ and $N=3$ electrons are shifted in phase.
For $N=3$ the transitions occur for the same magnetic fields as for $N=1$, i.e. at odd multiples of half of the flux quantum
and for $N=2$ the transitions occur at integer multiples of the flux quantum. For the ring with a single
defect the cusps related to the angular momentum transitions at high fields are replaced by continuous U or $\bigcap$-shaped
extrema for all $N$. For two symmetrically placed defects the spin-polarized ground-state of two electrons
is always of the odd parity [cf. Fig. \ref{spectra}(h)], while the ground-states of the single electron
as well as the ground-state for the spin polarized three electrons have an alternate parity when the magnetic field increases.
Hence in the chemical potential for two electrons we observe $V$ minima, related to the parity transitions of the single-electron
ground state, and $\bigcap$ maxima related to the smooth avoided crossing in the triplet part of the spectrum.
These avoided crossings for $N=2$ are translated into  $U$ minima in the $\mu_3$ plot. The maxima of the
chemical potential for three electrons are due to the parity transformations and are associated to the  $\Lambda$ shaped
maxima (for $N=3$ the spin polarization occurs for $B>4$ T and for $N=2$ for $B>3.5$ T).

Note that for both $N=2$ and $N=3$ the chemical potential in
elliptic and circular rings are the same at low magnetic fields. One
can distinguish between the symmetry of the confinement potential
only at high magnetic field when the Zeeman effect removes the
low-spin states of the ground-state energy level.

\begin{figure}[ht!]
\centerline{\hbox{\epsfysize=50mm \epsfbox[37 87 545 601]{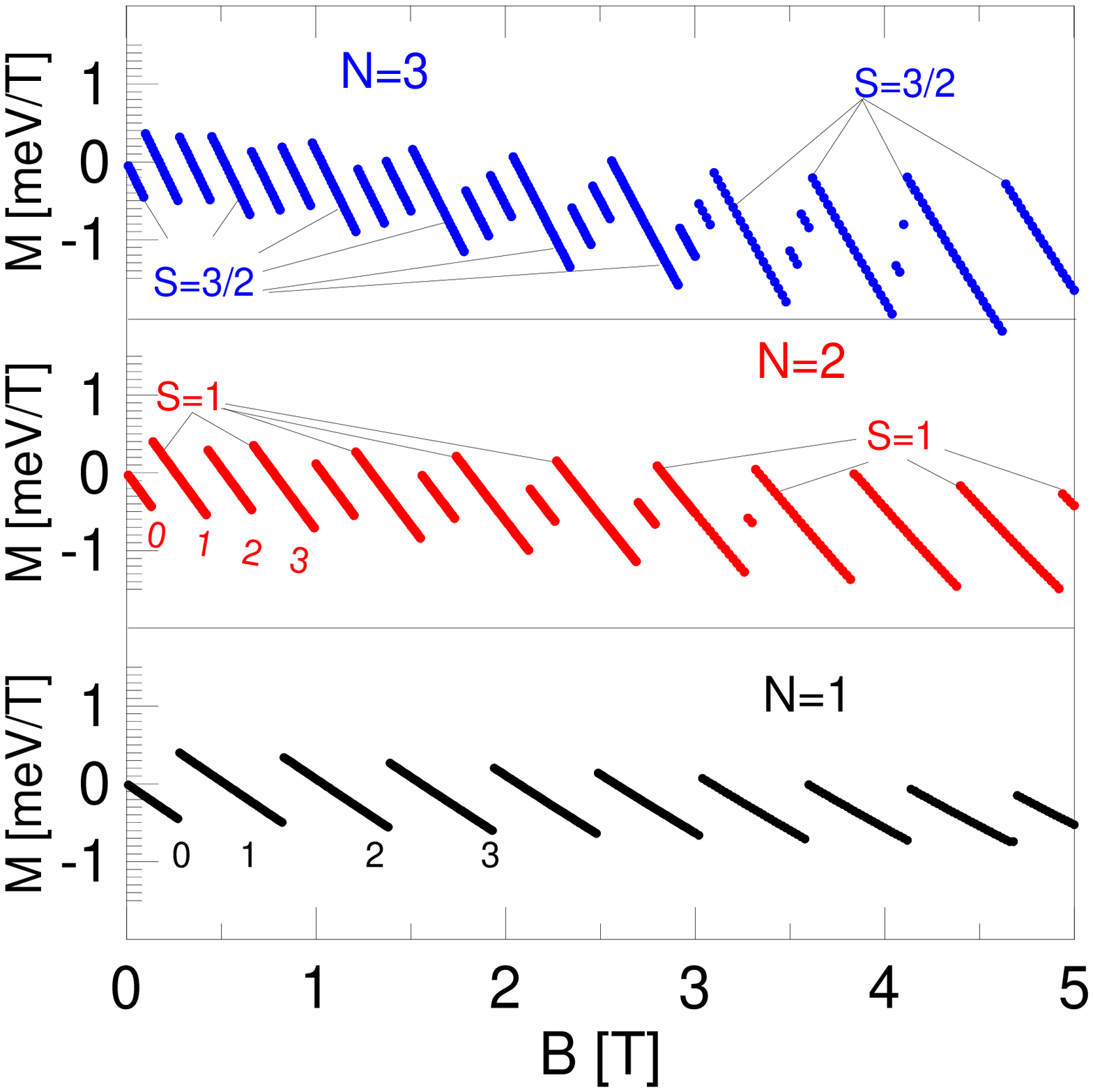} a)
               \epsfysize=50mm\epsfbox[37 87 545 601]{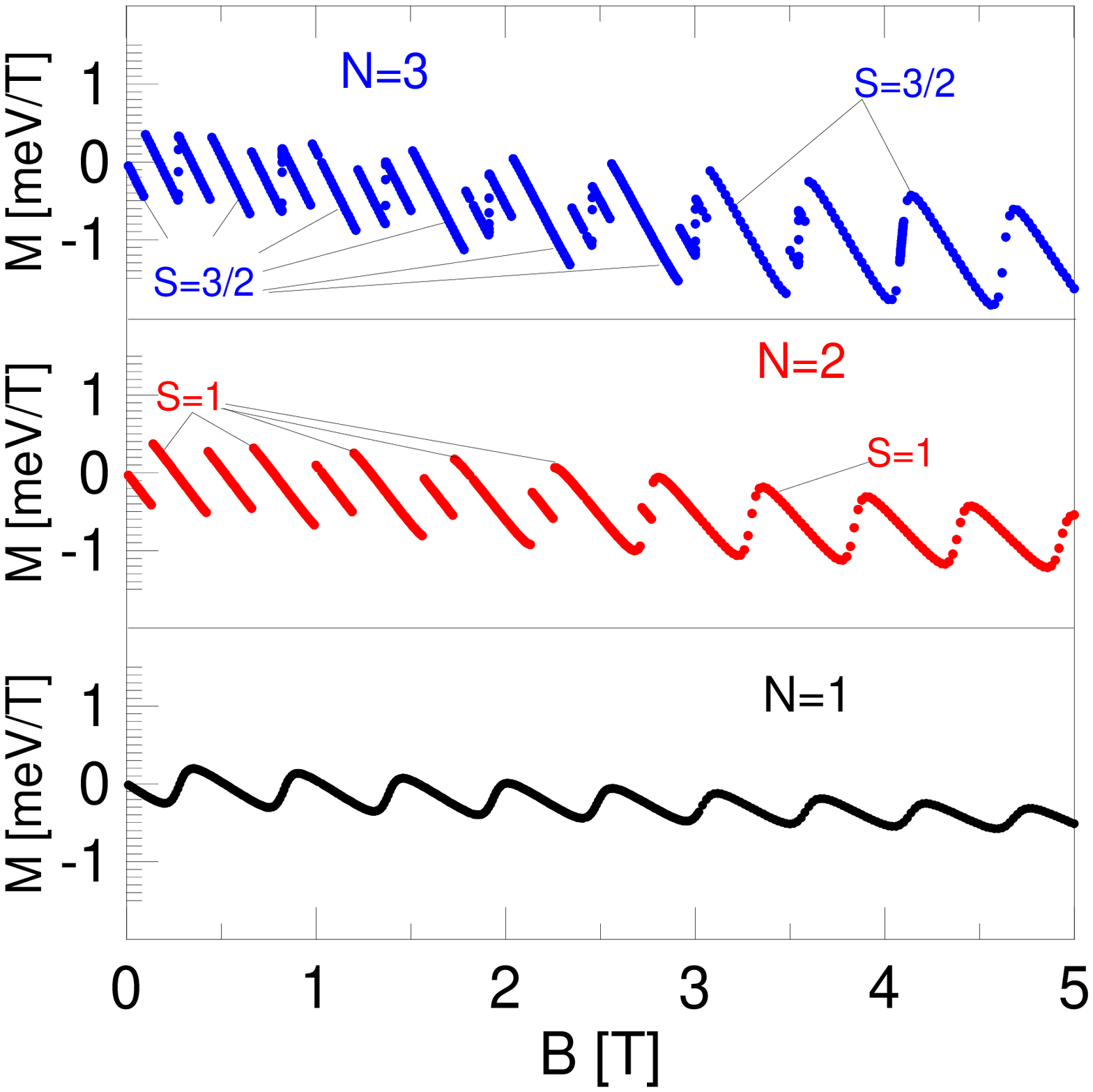}b) }
               \hbox{\epsfysize=50mm \epsfbox[37 87 545 601]{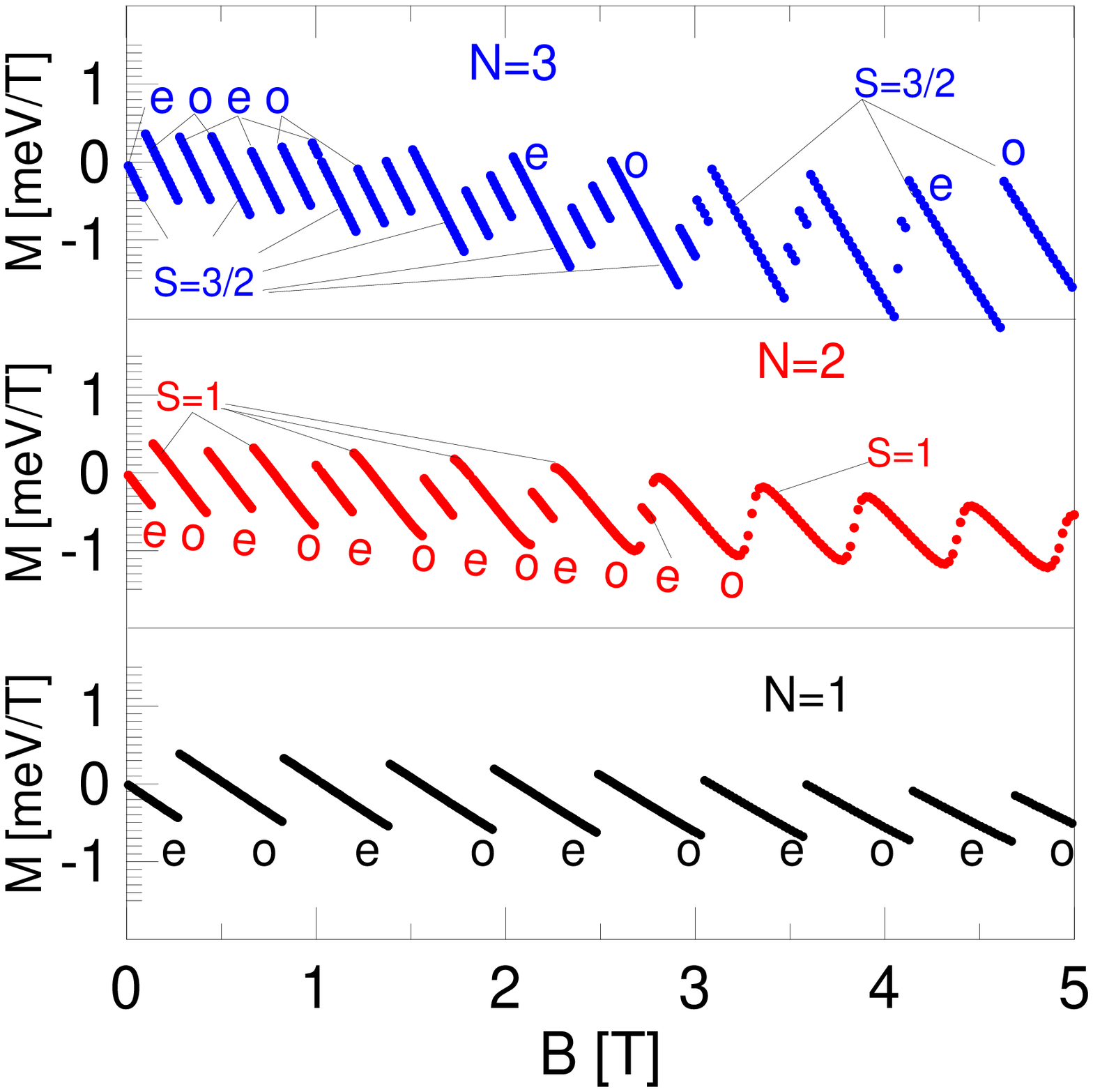} c)}}
\caption{Dipole moments for $N=1$ , 2 and 3 electrons account taken for the
spin Zeeman effect. In the plot for $N=2$ and $N=3$ electrons
the spin-polarized states are indicated.
Plot (a) corresponds to the clean ring, plot (b) to the ring with a single defect
and plot (c) to a ring with two defects. In (a) the ground-state angular momentum at $B=0$ is zero for any $N$,
and increases with each discontinuity (the $L$ values corresponding to the ground state
are listed for $N=1$ and $N=2$.) In (c) we marked the even and odd parity for all ground states
in the plot for $N=1$ with ''e`` and ''o`` letters. For $N=2$ and $N=3$ parity of some of the ground-states
is also given. The spin-polarized $N=2$ ground-state has the odd parity. }
\label{dipzjd}
\end{figure}

Fig. \ref{dipzjd} shows the dipole moment calculated for the ring with a single defect with the spin Zeeman
effect. In Fig. \ref{dipzjd}(a) we give the results for the clean ring, in Fig. \ref{dipzjd}(b) for
the single and in Fig. \ref{dipzjd}(c) for two defects.
In Fig. \ref{dipzjd}(a) the discontinuities are due to the angular momentum and spin transitions.
For $N=2$ and $N=3$ we notice enlargement of the magnetic fields in which the ground state is spin-polarized
as $B$ grows.
Finally, for high magnetic field  the ground state becomes permanently spin-polarized.
For the ring with a single defect the
discontinuities in the plot for $N=2$ and $N=3$ electrons are entirely due to the spin transitions.
At high field, when the low-spin states go up on the energy scale, the dipole moment becomes
a smooth functions of the magnetic field. For the ring with two defects the dipole
moment of the two-electrons at high field is a smooth function of $B$ (the ground state is of the odd parity),
while discontinuities related to the ground-state parity transitions are observed for both $N=1$ and $N=3$.
In the absence of the Zeeman effect the pattern of chemical potentials for $N=2$ and $N=3$ electrons
is the same for both the circular ring and the ring with symmetrically placed defects.
The qualitative difference between the two ring with a single or two defects is only observed for higher magnetic fields
when the Zeeman effect becomes important.

\section{Discussion}

The persistent currents circulating within the quantum ring produce
dipole moment whose orientations oscillates from parallel to
antiparallel to the external magnetic field as the latter increases.
The direction of the dipole moment results of the competition of
diamagnetic and paramagnetic contributions to the persistent
current. Both types of currents produce balanced contributions to
the dipole moment - canceling each other - for the magnetic fields
values corresponding to the vanishing derivative of the ground-state
energy with respect to the magnetic field.

The energy minima in function of the magnetic field have smooth
character for clean -- ideally circular -- quantum rings as well as
for the rings with defects, placed symmetrically or not with respect
to the origin. For circular rings in the energy minima the
diamagnetic and paramagnetic currents form concentric loops
circulating in opposite directions in the inner and outer edge of
the charge density. Both the loops generate magnetic field which is
oriented antiparallel to the external field inside the region
occupied by the charge density. For rings in which the circular
symmetry is perturbed by the defects the degree of the charge
density and persistent currents deviation from circular oscillates
with the magnetic field. These quantities are most circular and
resemble the solution for the ideally clean ring at the
local energy minima.

Smooth ground-state energy maxima which also correspond to vanishing dipole moment
occur only for quantum rings of lower than circular symmetry. They are due to avoided
crossings that are opened in the spectrum for mixing of angular momentum eigenstates.
They correspond to maximal deformation of the charge density
with one of electron most strongly localized at the attractive defect, and the other
electrons pinned by the Coulomb potential of the trapped electron.
The persistent current form loops around each single-electron charge density tending to produce the magnetic
field opposite to the external one within the region occupied by the electrons.

For the smooth energy maxima the character of the dipole moment undergoes a smooth transition from
diamagnetic to paramagnetic. For low magnetic fields and $N=1$ the smooth energy maxima occur in only
in the ring with a single defect. For $N=1$ and symmetrically placed defects
the even-odd symmetry transitions with a rapid reorientation of the currents from the paramagnetic to diamagnetic
occurs similarly as in the circular ring.  In low magnetic fields the smooth maxima occur in
the spin-polarized part of the two-electron spectrum for both a single and two defects.
For the spin-polarized three-electron system similar smooth maximum occurs only for the ring with a single defect
and not for two symmetrically placed defects. At higher magnetic field the smooth
energy maxima appear in the ground-state due to the spin Zeeman effect.
At the smooth energy maxima the charge density crystalizes in form of a Wigner
molecule that appears in the laboratory frame of reference. Note,
that whenever the Wigner crystallization occurs ($N=1, 2$ and 3 for
a single defect, and $N=2$ for two defects) the symmetry of the
configuration of the single-electron islands agrees with the
symmetry of the confinement potential. Moreover, the configuration
formed by the single-electron islands corresponds to a unique
lowest-energy classical configuration of point charges. In the other
studied cases of the circular ring for any $N$ as well as  the ring with two
symmetrically spaced defects and odd $N$ there are
more than one equivalent classical configuration. In the first case
there is an infinite number of classical configuration oriented at
an arbitrary angle for the clean quantum ring. For the ring with two symmetrically placed defects
there are two equivalent configurations with one or the other defect
occupied by an electron). For both the cases formation of
single-electron charge islands in a classical configurations would
break the symmetry of the confinement potential and thus it is not
realized in the non-degenerate ground-state. Absence of the smooth
energy maxima, i.e., absence of the Wigner crystallization of the
charges is accompanied by symmetry transitions in function of the
magnetic field. For circular rings this involves the angular
momentum transitions and for the rings of elliptical symmetry the
even-odd parity transformations.

The agreement of the symmetry of the classical point-charge
distribution and the symmetry of the confinement potential is a
necessary condition for the formation of the Wigner crystalized
charge density in the ground-state. Since the formation is related
to a smooth energy maximum we can indicate signatures of the Wigner
crystallization in the chemical potential spectrum as well as in the
dependence of the dipole moment generated by the persistent currents
in function of the magnetic field. In the chemical potential of $N$
electrons the $U$ shaped minima are related to the density pinning
of the $N-1$ electron system and the $\bigcap$ maxima to the
crystallization occurring in the $N$ electron system. The dipole
moment of the systems in which the Wigner crystallization occurs
becomes a continuous function of the magnetic field at high magnetic
field.


The magnetic field generated by the currents is weak -- of the order
of 100 nT for the external field of the order of 1 T. Although these values stay
within reach of commercial NMR Teslameters (that allow to detect
the field of the order of 10 T with the precision of 1 nT), a significant
interaction of electrons through the magnetic field they generate is
excluded. Note that this is not the case for mesoscopic metal rings
in which the magnetostatic electron-electron interactions can be
significant.\cite{zipper}

 \section{Summary and Conclusions}
 We studied the systems of up to three electrons in a circular quantum ring with a single or two symmetrically placed
 weak attractive defects in context of the Wigner crystallization of the confined electron charge using the configuration interaction
 method.
We showed that the pinning of the Wigner molecules in the laboratory
frame occurs only
 when the classical symmetry of the molecule
 agrees with the symmetry of the confinement potential.
 We showed that the pinning of the single-electron charge density maxima is associated
 with a local continuous  maximum of the ground-state energy as a function of the magnetic field.
 For systems in which the pinning is forbidden a cusp of the ground state energy due to a ground-state symmetry
 transformation is observed instead of the smooth maximum.
 The character of the maximum -- smooth or cusp-like --  affects the
 experimentally accessible properties of the system. In particular -- when the pinning is
 allowed by the symmetry -- the smooth maximum is
 associated with a continuous reversal of the dipole moment from
 diamagnetic to paramagnetic orientation,
 as well as with a presence of smooth
 extrema of the chemical potentials. For systems in which the pinning
 is forbidden by the symmetry the maxima of the ground-state
 energy are discontinuous which results in appearance
 of a discontinuous dipole moment reorientation and replacement of the
 smooth extrema of the chemical potentials by cusps.
     For $N>1$ electrons
 the maximal pinning of the charge densities in the ground-state energy
 appears only at high magnetic fields due to the spin Zeeman effect.

 We have shown that at the ground-state energy minima
 the charge density is the closest to circular.
 Then, the current has a form of two loops running in opposite direction at the inner and outer edges of an approximately circular electron density. Both the current loops tend to screen the external field within the confined charge density.
    Formation of most pronounced single-electron islands at the energy maxima is associated with  broken current loops around the ring
  and vanishing dipole moment of the system. The single-electron islands
have current vortices circulating at their edges.
  These currents have a diamagnetic character, and
generate magnetic field of opposite orientation to the external
field within the single-electron charge islands. The distribution of
the charge density islands is therefore visible in the maps of the
magnetic field generated by the persistent currents.

{\bf Acknowledgments}
This work was partly supported by the Polish Ministry for Science
  and Higher Education and by the EU Network of Excellence: SANDiE.


\begin{thebibliography}{00}
\bibitem{lorke} A. Lorke, R.J. Luyken, A.O. Govorov, J.P. Kotthaus, J. M. Garcia
and P. M. Petroff, Phys. Rev. Lett. {\bf 84}, 2223 (2000).
\bibitem{so} W. G. van der Wiel, Yu.V. Nazarov, S. De Franceschi,
T. Fujisawa, J.M. Elzerman, E.W.G.M. Huizeling, S. Tarucha, and L.P.
Kouwenhoven, Phys. Rev. B {\bf 67}, 033307 (2003).
\bibitem{hawrylak}
M. Bayer, M. Korkusinski, P. Hawrylak, T. Gutbrod, M. Michel  and A.
Forchel, Phys. Rev. Lett. {\bf 90} 186801 (2003).
\bibitem{buti} M. B\"uttiker, Y. Imry, and M. Y. Azbel, Phys. Rev. A {\bf 30}, 1982
(1984). \bibitem{transport} A. Fuhrer, S. L\"uscher, T. Ihn, T. Heinzel, K. Ensslin, W. Wegscheider, and M. Bichler, Nature (London), 413, 822 (2001).
\bibitem{rev} S. Viefers, P. Koskinen, P. Singa Deo, and M. Manninen, Physica
E (Amsterdam) {\bf 21}, 1 (2004).
\bibitem{fomin}N.A.J.M. Kleemans, I.M.A. Bominaar-Silkens, V.M.
Fomin, V.N. Gladilin, D. Granados, A.G. Taboada, J.M. Garcia, P.
Offermans, U. Zeitler, P.C.M. Christianen, J.C. Maan, J.T. Devreese,
and P.M. Koenraad, Phys. Rev. Lett. {\bf 99}, 146808 (2007).
\bibitem{mm2s}
D. Mailly, C. Chapelier, and A. Benoit,  {Phys. Rev. Lett.} {\bf
70}, {2020} (1993).
\bibitem{foomin} V. M. Fomin, V. N. Gladilin,  J. T. Devreese, N. A. J. M. Kleemans and P. M.
Koenraad, Phys. Rev. B {\bf 77}, 205326 (2008).
\bibitem{lqd} K. Jauregui, W. H\"ausler, and B. Kramer, Europhys. Lett. {\bf 24}, 581
(1993); W. H\"ausler and B. Kramer, Phys. Rev. B {\bf 47}, 16 353 (1993).
\bibitem{mani} S.M. Reiman and M. Manninen, Rev. Mod. Phys. {\bf 74}, 1283 (2002).
\bibitem{jain} C. Shi, G.S. Jeon and J.K. Jain, Phys. Rev. B {\bf 75}, 165302 (2007).
\bibitem{tavernier} M. B. Tavernier, E. Anisimovas, F. M. Peeters, B. Szafran, J. Adamowski, and S. Bednarek,
Phys. Rev. B {\bf 68}, 205305 (2003).
\bibitem{sigmund} M. Siegmund, M. Hofmann, and O. Pankratov, arXiv: 0711.2937.
\bibitem{szafranchwiej} B. Szafran, F. M. Peeters, S. Bednarek, T. Chwiej, and J. Adamowski, Phys. Rev. B {\bf 70} 035401 (2004).
\bibitem{glazman} L.I. Glazman, I.M. Ruzin, and B. I. Shklovskii, Phys. Rev. B {\bf 45}, 8454 (1992).
\bibitem{maksym} P.A. Maksym, H. Immamura, G.P. Mallon, and H. Aoki, J. Phys.:
Condens. Matter {\bf12}, R299 (2000).
\bibitem{peeters} V.M. Bedanov and F.M. Peeters, Phys. Rev. B {\bf 49}, 2667 (1994).
\bibitem{pcct4} V.M. Fomin, V.N. Gladilin, S.N. Klimin, J.T.
Devreese, N.A.J.M. Kleemans, and P.M. Koenraad, Phys. Rev. B {\bf
76}, 235320 (2007).
\bibitem{pcct0} T. Chakraborty and P. Pietil\"{a}inen,  {Phys. Rev. B}
{\bf 50}, {8460}  {(1994)}; A.O. Govorov, S.E. Ulloa, K. Karrai, and
R.J. Warburton, {Phys. Rev. B} {\bf 66}, {081309(R)}  {(2002)}; J.I.
Climente, J. Planelles, and J.L. Movilla,
 {Phys. Rev. B} {\bf 70}, {081301(R)} {(2004)}; Y.V. Pershin and C.
Piermarocchi,  {Phys. Rev. B} {\bf 72}, {125348}  {(2005)}.
\bibitem{low} P-O. Lowdin, Rev. Mod. Phys. {\bf 36}, 966 (1964).
\bibitem{fqhe}  P. Pietil\"ainen, P. Hyv\"onen, and T. Chakraborty,
Europhys. Lett. {\bf 36}, 533 (1996).
\bibitem{ostatni} J.I. Climente and J Planelles, J. Phys.: Condens.
Matter {\bf 20}, 035212 (2008).
\bibitem{odnosnikd} Note for the single-electron angular momentum eigenstate
with wave function $\psi$ in a circular ring the angular paramagnetic current is given by
${\bf j}_p(r)|_\phi=-\frac{e\hbar}{ m r} l |\psi|^2$, while the diamagnetic current
is expressed by ${\bf j}_d(r)|_\phi=\frac{-e^2}{2m} B r |\psi|^2$. Therefore, the paramagnetic current
depends on the magnetic field only through the wave function, and the contribution to the
diamagnetic current is larger at the outer edge of the ring. For $B<0$ (as applied here) the ground state corresponds
to nonnegative $l$, hence the opposite orientations of the currents.
\bibitem{naszarx} T. Chwiej and B. Szafran,	Phys. Rev. B {\bf 78}, 245306 (2008).
\bibitem{kainz} J. Kainz, S.A. Mikhailov, A. Wensauer, and U. R\"ossler, Phys.
Rev. B {\bf 65}, 115305 (2002).
\bibitem{szafran}B. Szafran,  F.M. Peeters, and  S. Bednarek, Phys. Rev. B  {\bf 70}, 205318 (2004).
\bibitem{yann} C. Yannouleas, Rep. Prog. Phys. {\bf 70}, 2067 (2007).
\bibitem{szafranes} B. Szafran, Phys. Rev. B {\bf 77}, 205313 (2008).
\bibitem{zipper} M. Lisowski, E. Zipper, and M. Stebelski, Phys. Rev. B {\bf 59}, 8305 (1999).
\end{thebibliography}
\end{document}